\documentclass[epj]{svjour}
\usepackage{graphicx}
\usepackage{subfigure}
\usepackage{marvosym}
\usepackage{amsmath}
\usepackage{amssymb}
\usepackage{multirow}
\usepackage{fancyhdr} 
\usepackage[normalem]{ulem}
\newcommand{\pt}{$p_T$}

\newcommand{\f}{$\phi$}

\newcommand{\gevcs}{\textrm{GeV}}
\newcommand{\gevc}{\textrm{GeV}}

\let\oldmarginpar\marginpar
\renewcommand\marginpar[1]{\-\oldmarginpar[\raggedleft\footnotesize #1]%
{\raggedright\footnotesize #1}}

\begin{document}
\title{$\phi$ Production in In-In Collisions at 158~A\,GeV}
%
\author{
  R.~Arnaldi\inst{12},
  K.~Banicz\inst{5,7},
  K.~Borer\inst{1},
  J.~Castor\inst{6},
  B.~Chaurand\inst{9},
  W.~Chen\inst{2},
  C.~Cical\`o\inst{4},
  A.~Colla\inst{11,12},
  P.~Cortese\inst{11,12},
  S.~Damjanovic\inst{5,7},
  A.~David\inst{5,8},
  A.~de~Falco\inst{3,4},
  A.~Devaux\inst{6},
  L.~Ducroux\inst{13},
  H.~En'yo\inst{10},
  J.~Fargeix\inst{6},
  A.~Ferretti\inst{11,12},
  M.~Floris\inst{3,4},
  A.~F\"orster\inst{5},
  P.~Force\inst{6},
  N.~Guettet\inst{5,6},
  A.~Guichard\inst{13},
  H.~Gulkanyan\inst{14},
  J.~Heuser\inst{10},
  M.~Keil\inst{5,8},
  L.~Kluberg\inst{5,9},
  Z.~Li\inst{2},
  C.~Louren\c{c}o\inst{5},
  J.~Lozano\inst{8},
  F.~Manso\inst{6},
  P.~Martins\inst{5,8},
  A.~Masoni\inst{4},
  A.~Neves\inst{8},
  H.~Ohnishi\inst{10},
  C.~Oppedisano\inst{12},
  P.~Parracho\inst{5,8},
  P.~Pillot\inst{13},
  T.~Poghosyan\inst{14},
  G.~Puddu\inst{3,4},
  E.~Radermacher\inst{5},
  P.~Ramalhete\inst{5,8},
  P.~Rosinsky\inst{5},
  E.~Scomparin\inst{12},
  J.~Seixas\inst{8},
  S.~Serci\inst{3,4},
  R.~Shahoyan\inst{5,8},
  P.~Sonderegger\inst{8},
  H.J.~Specht\inst{7},
  R.~Tieulent\inst{13},
  A.~Uras\inst{3,4},
  G.~Usai\inst{3,4},
  R.~Veenhof\inst{8} \and
  H.K.~W\"ohri\inst{4,8}\\
  (NA60 Collaboration)}
%
\institute{Laboratory for High Energy Physics, Bern, Switzerland;
  \and BNL, Upton, NY, USA;
  \and Universit\`a di Cagliari, Italy;
  \and INFN Cagliari, Italy;
  \and CERN, Geneva, Switzerland;
  \and Universit\'e  Blaise Pascal and CNRS-IN2P3, Clermont-Ferrand, France
  \and Physikalisches Institut der Universit\"at Heidelberg, Germany
  \and Istituto Superior T\'ecnico, Lisbon, Portugal
  \and LLR, Ecole Polytechnique and CNRS-IN2P3, Palaiseau, France;
  \and RIKEN, Wako, Saitama, Japan;
  \and Universit\`a di Torino, Italy;
  \and INFN Torino, Italy;
  \and IPNL, Universit\'e de Lyon, Universit\'e Lyon 1, CNRS/IN2P3, Villeurbanne, France
  \and YerPhI, Yerevan Physics Institute, Yerevan, Armenia}
\date{Received: date / Revised version: date}
%
\abstract{The NA60 experiment has measured muon pair production in
  In-In collisions at 158 AGeV at the CERN SPS.  This paper presents a
  high statistics measurement of $\phi\to\mu\mu$ meson
  production. Differential spectra, yields, mass and width are
  measured as a function of centrality and compared to previous
  measurements in other colliding systems at the same energy. The
  width of the rapidity distribution is found to be constant as a
  function of centrality, compatible with previous results. The decay
  muon polar angle distribution is measured in several reference
  frames. No evidence of polarization is found as a function of
  transverse momentum and centrality. The analysis of the $p_{T}$
  spectra shows that the $\phi$ has a small radial flow, implying a
  weak coupling to the medium.  The $T_{eff}$ parameter measured in
  In-In collisions suggests that the high value observed in Pb-Pb in
  the kaon channel is difficult to reconcile with radial flow
  alone. The absolute yield is compared to results in Pb-Pb
  collisions: though significantly smaller than measured by NA50 in
  the muon channel, it is found to exceed the NA49 and CERES data in
  the kaon channel at any centrality. The mass and width are found to
  be compatible with the PDG values at any centrality and at any
  $p_{T}$: no evidence for in-medium modifications is observed.
} 
\authorrunning{R. Arnaldi \emph{et al}}
\maketitle

\section{Introduction}
\label{sec:f-production-heavy}
QCD predicts the occurrence of a phase transition from ha\-dro\-nic
matter to a deconfined plasma of quarks and gluons, when sufficiently
high energy densities are reached.  Strangeness enhancement was
proposed long ago as a signature of this phase
transition~\cite{Rafelski:1982pu,Shor:1984ui}. The \f~meson is a key
experimental probe in this context: due to its $s\bar{s}$ valence
quark content it permits to measure strangeness production. It was
suggested that, being a hidden strangeness state, the \f~should not be
sensitive to canonical suppression, and can thus be used to test
strangeness enhancement
models~\cite{Becattini:1997ii,Andronic:2003zv,Becattini:2005xt,Becattini:2008yn,Abelev:2008zk}.

Moreover, it has been argued that the spectral function of the \f~or
of its decay products could be modified in the medium.  This would
reflect in changes of its mass and partial decay widths.  Mass shifts
of vector mesons, as a consequence of the change of the quark
condensate in the medium, were originally proposed by Brown and
Rho~\cite{Brown:1991kk}.  Subsequently, the $\phi$ meson mass
dependence on the medium density was calculated with QCD sum
rules~\cite{Hatsuda:1991ez}. Modifications of the width in the context
of this dropping mass scenario were investigated, for instance, in the
model of Ref.~\cite{Pal:2002aw}. As an alternative approach, both the
mass and width changes were studied in the framework of hadronic many
body models (see \cite{vanHees:2007th} for an up-to-date discussion),
which predict a broadening and no or small mass shift. Finally, it was
suggested that in-medium changes could also appear in cold nuclear
matter~\cite{Cabrera:2002hc,Oset:2000eg}.

It is important to point out that, contrary to the $\rho$ meson which
has a lifetime of 1.3~fm, much smaller than the lifetime of the
fireball (7-10~fm), the $\phi$ has a lifetime of 44~fm.  Therefore
in-medium effects on the mass or width might be difficult to isolate
experimentally, and might only appear at very low $p_T$. A possible
way of tracing down those modifications is the comparison of the
particle yields and spectra in the dilepton and in the kaon channel:
since the $\phi$ is almost at the threshold for the $\phi\to K\bar{K}$
channel, even a small change in its spectral properties should lead to
a significant difference in the two channels~\cite{Shuryak:1999zh}. A
discrepancy in the two channels could also be ascribed to final state
effects related to in medium kaon absorption or
rescattering~\cite{Johnson:1999fv,Santini:2006cm}.

\begin{figure*}[t]
  \centering
  \includegraphics[width=1.0\textwidth]{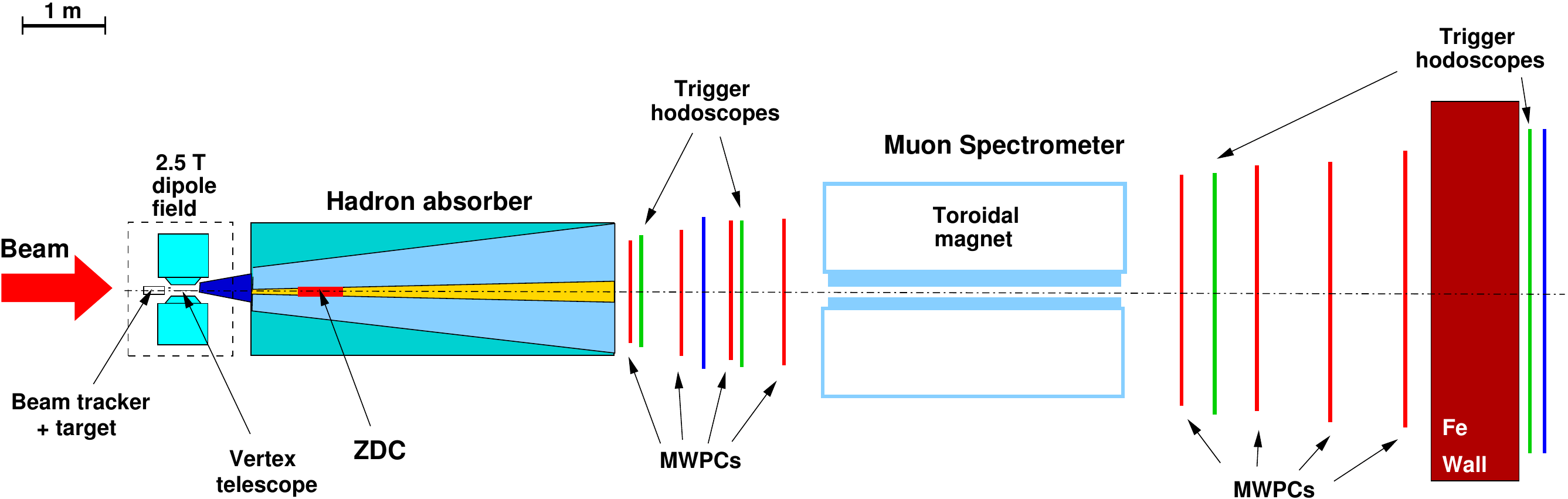}
  \caption{The NA60 experimental apparatus.}
  \label{fig:Apparatus1}
\end{figure*}
\begin{figure*}[t]
  \centering
  \vspace{0.9cm}
 \includegraphics[width=0.8\textwidth,height=4.5cm]{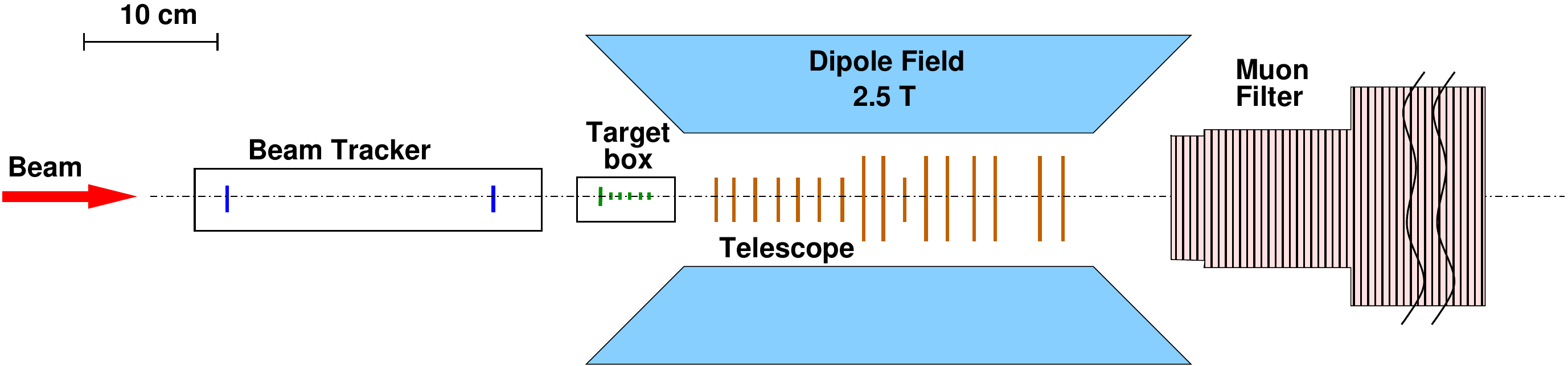}
  \vspace{0.8cm}
  \caption{Detail of the vertex detectors layout.}
  \label{fig:Apparatus2}
\end{figure*}

The KEK-PS E325 experiment reported evidence for a large excess to the
low mass side of $\phi$ peak in p-Cu at 12 GeV for $\beta \gamma <
1.25$, corresponding to slow $\phi$ mesons, having a higher chance to
decay inside the Cu nucleus (the effect was not observed in p-C and
for higher values of $\beta \gamma$)~\cite{Muto:2005za}.  This excess
was explained in terms of an in-medium modified component with mass
decreased by $\sim 35$~MeV and width broadened by a factor 3.6 with
respect to the PDG value. The reported yield associated to this excess
is very significant: 20\% of the total $\phi$ yield.

\f~production was studied extensively at the SPS by several
experiments.  NA49 measured \f~production in p-p, C-C, Si-Si and Pb-Pb
at 158~AGeV in the $\phi \to K^+K^-$
channel~\cite{Friese:2002re,Alt:2004wc,Afanasev:2000uu}. These data
show that the \f~yield breaks $N_{part}$ scaling and that the
\f~enhancement, quantified as $\left<\phi\right>/N_{part}$, saturates
already for Si-Si central collisions.  Pb-Pb collisions were also
studied in different experimental conditions by the NA50 experiment,
which found contrasting results. NA50 studied the $\phi \to \mu \mu$
channel~\cite{Alessandro:2003gy} with acceptance limited to $p_T >
1.1~\mathrm{GeV}$, while the NA49 measurement in the $\phi \to
K^+K^-$ channel was limited to $p_T < 1.6~\mathrm{GeV}$ because of
statistics.  Both experiments observed an enhancement of the \f~yield
with the size of the collision system.  The absolute yields, however,
disagree by a factor $\sim4$ in full phase
space~\cite{Rohrich:2001qi}. A recent NA50 analysis of a new sample
confirmed previous findings~\cite{JOU08-Phi}.  The NA49 and NA50
results disagree also on the $T_{eff}$ parameters extracted from the
transverse mass spectra fitted with the thermal ansatz. The NA49
$T_{eff}$ values are consistently larger and show a stronger
centrality dependence than those measured by NA50.  While this could
be partly due to the radial expansion, which flattens the
distributions at low $m_T$, the effect is too large to be explained by
radial flow alone.

Naively, kaon absorption or rescattering could prevent the
reconstruction of in-matter $\phi\to K\bar K$ decays, in particular at
low transverse momentum, while the $\phi$ mesons decaying in the
lepton channel would not be affected.  This would lead to reduced
yields and enhanced $T_{eff}$ in the $K\bar{K}$ channel as compared to the
$\mu\mu$ channel. Meson modification in the medium could also lead to
differences in the two channels, as mentioned above.  The difference
between NA49 and NA50, however, is larger than predicted by models
including all these
effects~\cite{Pal:2002aw,Shuryak:1999zh,Johnson:1999fv,Santini:2006cm,Holt:2004tp}.
The discrepancies in absolute yields and $T_{eff}$ became known as the
\emph{\f~puzzle}.

Production of \f~in central Pb-Pb collisions was recently studied also
by CERES~\cite{Adamova:2005jr} both in the kaon and dielectron
channels.  The measured $T_{eff}$ and yield are compatible with the
NA49 results.  However, due to the large uncertainty in the dielectron
channel, the authors could not rule out completely a possible
difference in the two channels, though significantly smaller than the
NA49/NA50 difference: the maximum possible enhancement in the leptonic
over hadronic channel was estimated to be 1.6 at 95\% confidence level.

In this paper we report on new measurements in the muon channel done
by the NA60 collaboration in In-In collisions at 158 AGeV incident
beam energy, which can help to clarify \f~production in nuclear
collisions. Transverse momentum and yields have been studied as a
function of centrality.  Transverse momentum spectra, measured in the
range $0-3~\mathrm{GeV}$, and yields are compared to previous
results, providing further insight on the \f~puzzle.  The mass and
width of the \f~are studied as a function of centrality and $p_T$, in
order to search for possible in-medium effects. The results are
compared to previous measurements and existing theoretical predictions
for In-In collisions.
In addition, the rapidity and the decay angular distributions in three
reference frames (Collins-Soper, Gottfried-Jackson and the helicity
frame) are also measured.  In elementary collisions, the \f~decay
angular distributions can convey information on the production
mechanism~\cite{Ellis:1994ww}. No measurements exist in p-p, while the
ACCMOR collaboration performed measurements of the \f~polarization in
pion, proton and kaon - Be
interactions~\cite{Dijkstra:1986my,Dijkstra:1986na} . A small but
significant transverse polarization was reported at low $p_T$ in the
Gottfired-Jackson frame, which was related to the parton fusion model
to estimate the average mass of the fusing partons.  More recently,
NA60 measured the angular distributions in In-In collisions,
integrated in centrality and for $p_T>0.6~\mathrm{GeV}$, for the
excess dimuons below 1~GeV (dominated by
$\pi^+\pi^-\to\rho\to\mu^+\mu^-$) and the vector mesons $\omega$ and
$\phi$ ~\cite{Arnaldi:2008gp}.  The reported absence of any anisotropy
indicates the completely random orientation of the interacting hadrons
in 3 dimensions as expected in the case of a thermalized medium.  In
this paper we extend these results by performing the study as a
function of centrality and $p_T$.

\section{The NA60 apparatus}
\label{sec:na60-experiment}

In this section the detectors of the NA60 apparatus used in the
analysis are briefly described. A more detailed description can be
found in Ref.~\cite{Usai:2005zh}.

The 17-m-long muon spectrometer, previously used by the NA38 and NA50
experiments~\cite{Abreu:1997ji}, is complemented by new silicon
detectors in the vertex region.  The experimental apparatus is shown
in Fig.~\ref{fig:Apparatus1}. The muon spectrometer is composed of 8
multi-wire proportional chambers and 4 scintillator trigger
hodoscopes, divided into a ``forward'' and a ``backward'' telescope by
the 5-m-long ACM (Air Core Magnet) toroidal magnet.  A 120 cm thick
iron wall, located just before the last trigger hodoscope and after
the tracking chambers, ensures that only muons can trigger the
apparatus, without degradation of the tracking accuracy.  The angular
acceptance is $35<\theta<120$ mrad; the rapidity coverage for the
\f~is $3\lesssim y \lesssim 4.2$.

A 12 $\lambda_i$ hadron absorber, mostly made of graphite, is placed
in front of the tracking system of the muon spectrometer, as close as
possible to the target to minimize the fraction of hadrons that decay
into muons.  The identification provided by this ``muon filter'' leads
to a very selective trigger, enabling the experiment to run at very
high luminosities.  However, the material in the hadron absorber also
degrades the kinematics of the muons, because of energy loss
fluctuations and multiple scattering. This affects in particular the
angular resolution, which is $\sim10$~mrad for typical muons from
$\phi$ decays.

\begin{figure}[ptb]
  \centering
  \includegraphics[width=0.4\textwidth]{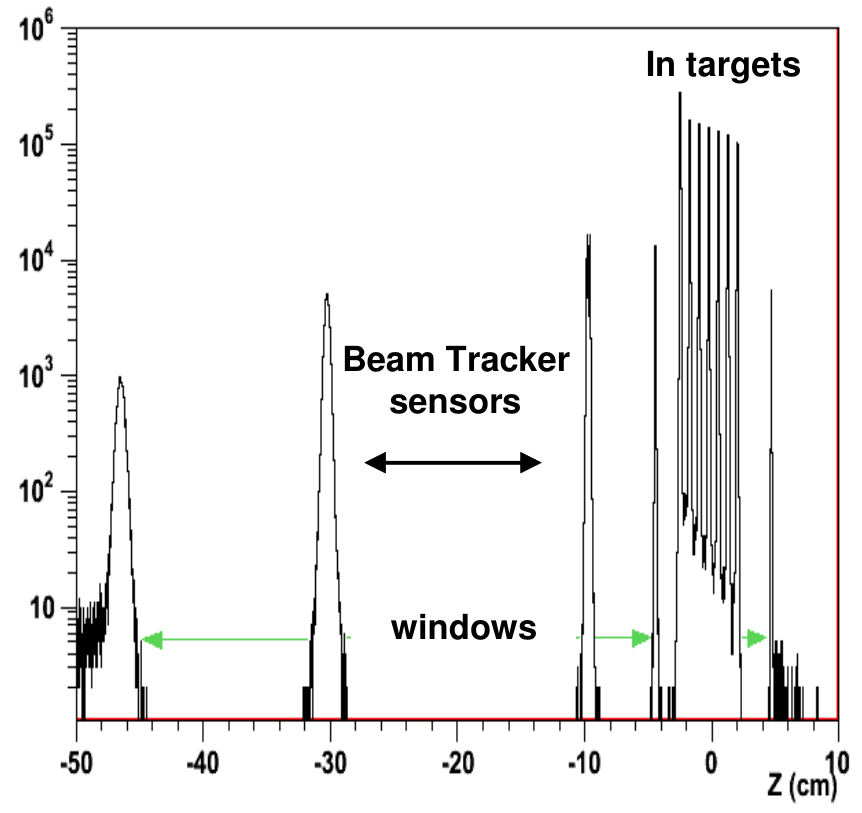}
  \caption{Longitudinal coordinate of interaction point.}
  \label{fig:Zvertex}
\end{figure}

The layout of the vertex region is shown schematically in
Fig.~\ref{fig:Apparatus2}.  Downstream of the target and before the
hadron absorber, a silicon tracking telescope~\cite{Banicz:2004mf}
embedded inside a 2.5 T magnetic dipole field measures the charged
tracks produced in the interaction.  The muons in the vertex region
are then identified by matching them to the tracks reconstructed in
the muon spectrometer. As described in the next section, this
significantly improves the performance of the detector.  The telescope
used during the indium run consisted of 16 independent detector
planes, providing 11 tracking points, arranged along the beam axis
over a length of $\sim26$ cm, starting at $\sim7$ cm from the target
centre.  The telescope covers the angular acceptance of the muon
spectrometer, with the first planes smaller than the last ones.  Each
plane consists of several single-chip pixel detector assemblies, 96 in
total, which are mounted on a planar ceramic support. The assemblies
are made of radiation tolerant, $750~\mu$m thick ALICE1LHCb pixel
readout chips~\cite{Wyllie:1999uv}, bump-bonded to $300~\mu$m thick
p-on-n silicon pixel sensors, with $425 \times 50~\mu$m$^2$
pixels. Each of the single chip assemblies is a matrix of 32 columns
and 256 rows. They have been shown to remain fully functional up to a
radiation dose of 12 Mrad~\cite{vanHunen:2001kc}. The detector coped
well with the radiation level of $\sim$1 Mrad per week, reached in the
cells around the beam axis of the first planes in the NA60
environment~\cite{Keil:2005zq}. The telescope tracks all charged
particles produced in the collision, thus providing an estimate of the
collision centrality.

Incoming ions are tracked by a silicon microstrip beam tracker,
composed of two stations placed 20~cm apart (30~cm and 10~cm upstream
of the target).  The 400~\(\mu \mathrm{m}\) thick sensors have 24
strips of 50~$\mu$m pitch, allowing to infer the transverse
coordinates of the interaction point, at the target, with a resolution
of 20~$\mu$m~\cite{Rosinsky:2003dd}.  They are operated at cryogenic
temperatures (130~K), to remain sensitive to the beam ions even after
collecting considerable radiation doses~\cite{Palmieri:1998zj}.  The
``back planes'' of the four sensors extend over a larger area and
constitute a very good beam intensity counter. 

Embedded inside the hadron absorber, a Zero Degree Calorimeter (ZDC)
provides an alternative estimate of the collision centrality by
measuring the energy carried by the projectile nucleons which have not
taken part in the interaction
(spectators)~\cite{Arnaldi:1998fk,colla}. The ZDC and the beam tracker
were used to measure the beam intensity in the absolute yield
measurement (sec.~\ref{sec:yield}).

\begin{figure*}[ptb]
  \centering
  \includegraphics[width=0.48\textwidth,height=0.48\textwidth]{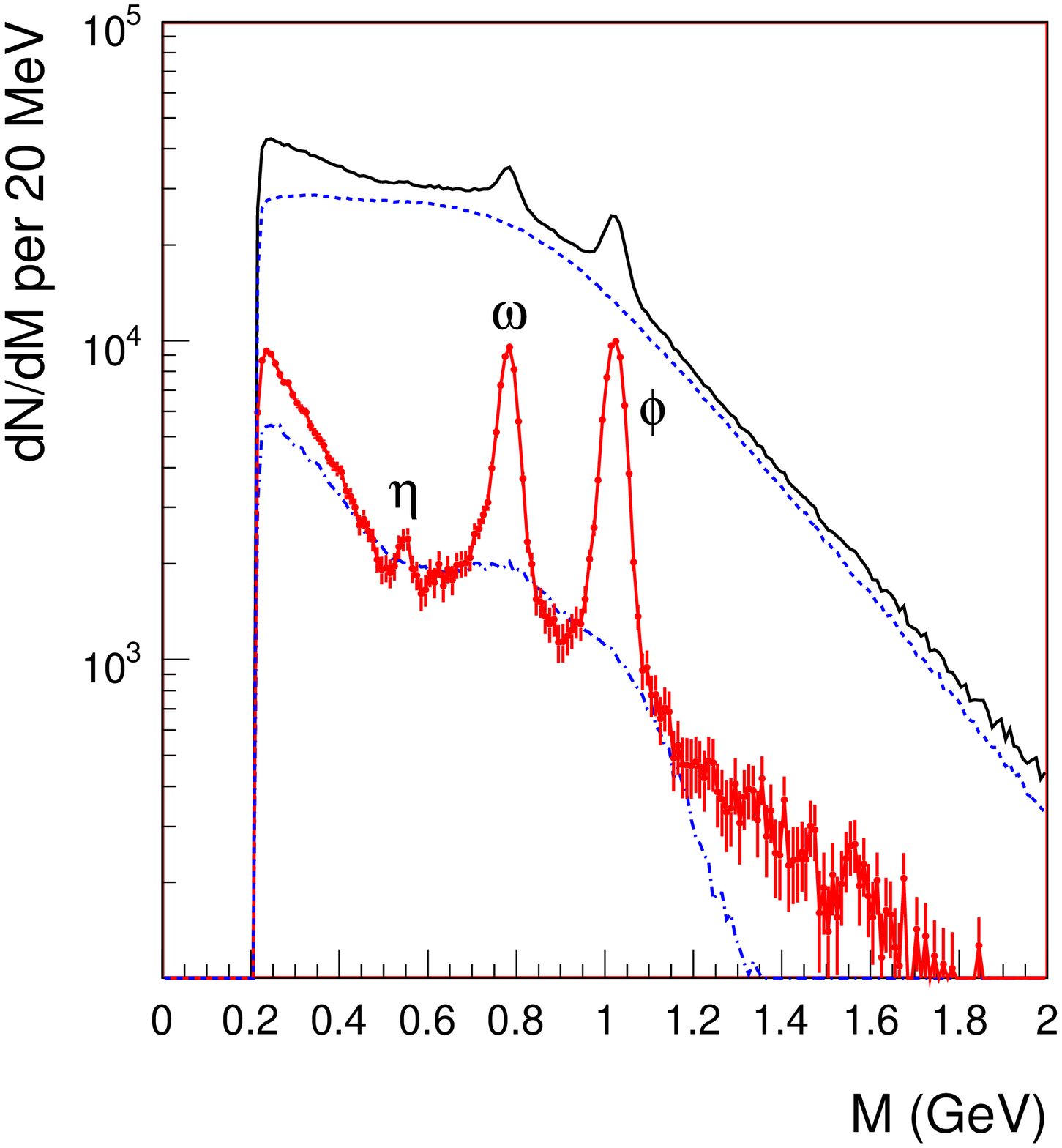}
  \includegraphics[width=0.48\textwidth,height=0.48\textwidth]{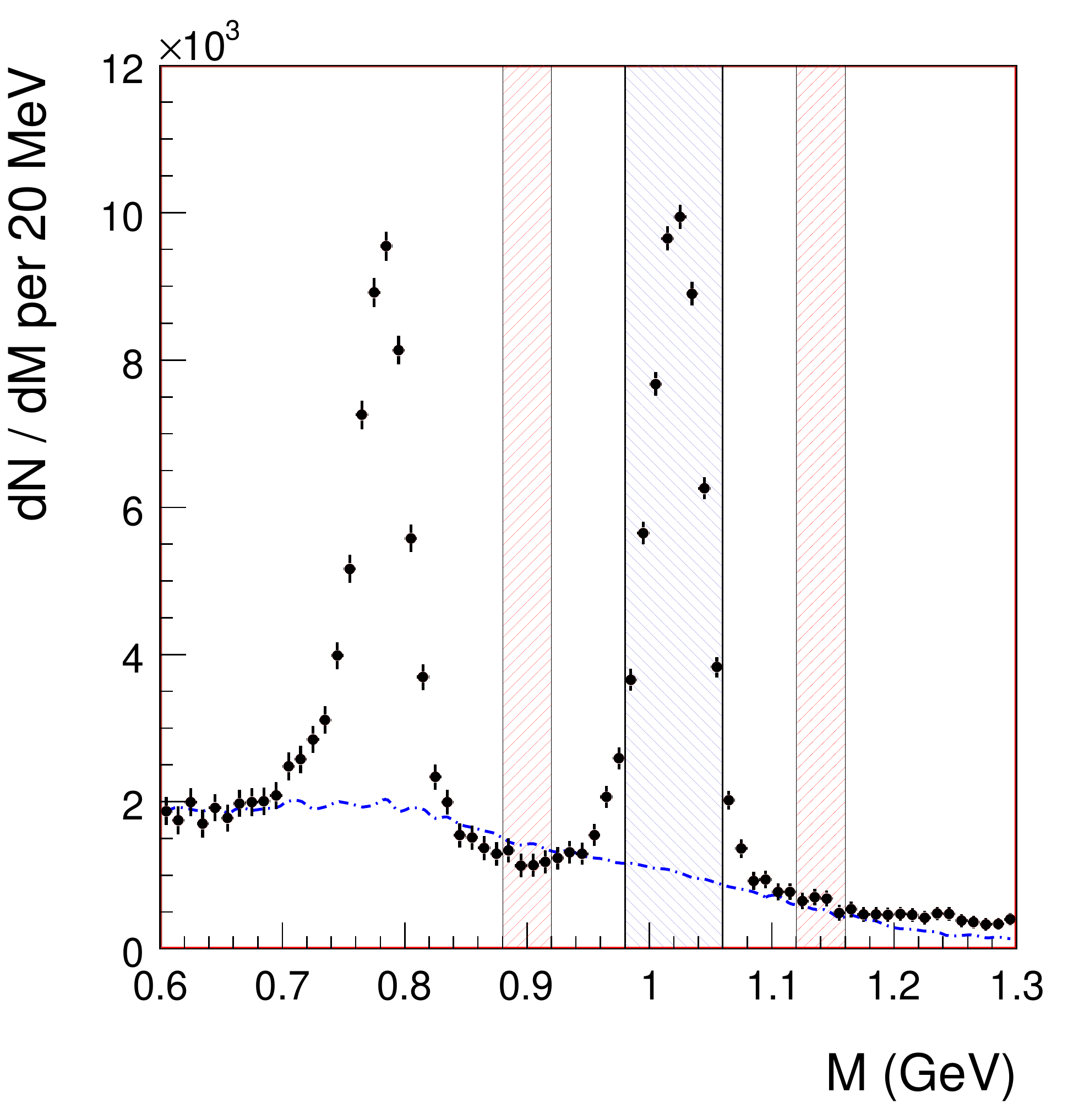}
  \caption{Left panel: raw mass distribution (continuous line),
    combinatorial background (dashed line), fake tracks (dot-dashed
    line) and mass distribution after combinatorial background and fake
    tracks subtraction (continuous curve with error bars). Right panel:
    mass region around the $\phi$ peak and mass windows used in the
    analysis after backgrounds subtraction. The dot-dashed distribution
    is the total fake matches distribution.}
  \label{fig:clean_mass}
\end{figure*}

\begin{figure*}[ptb]
  \centering
  \subfigure{\includegraphics[width=0.48\textwidth,height=0.48\textwidth]{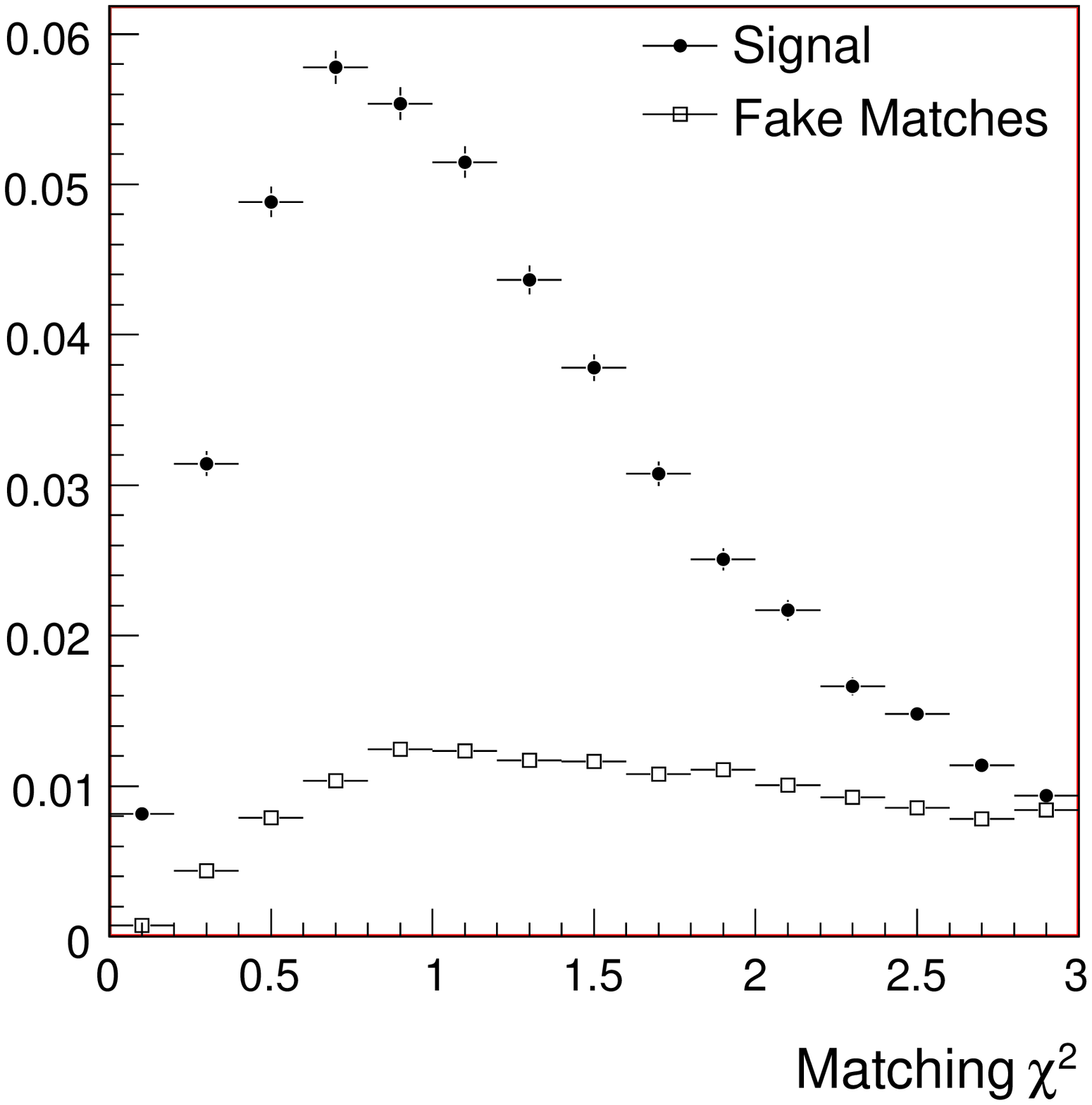}}
  \subfigure{\includegraphics[width=0.48\textwidth,height=0.48\textwidth]{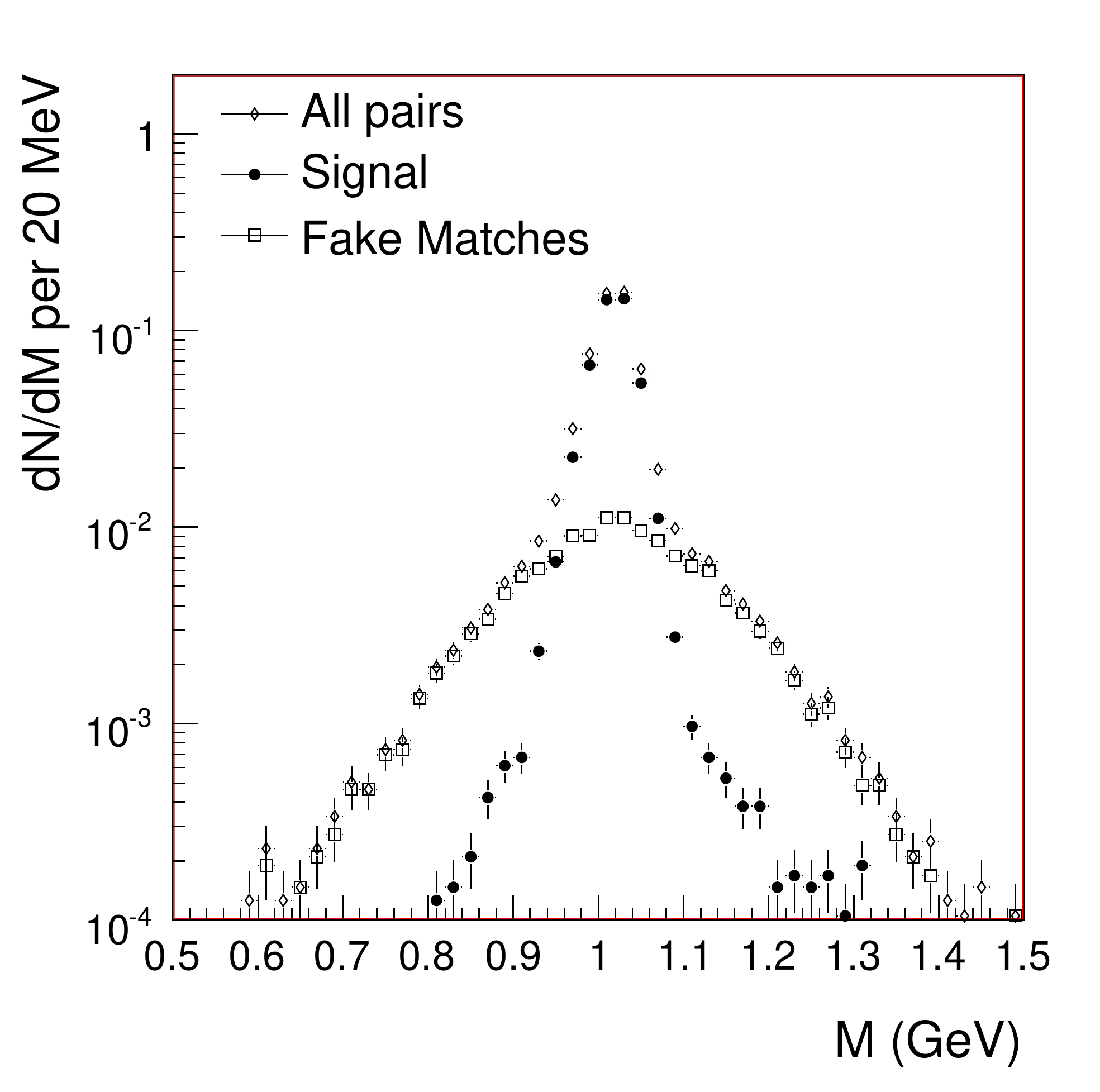}}

  \caption{Matching $\chi^2$ (left) and mass distribution (right) of
    signal and fake matches for simulated $\phi$ events.}
  \label{fig:MatchingChi2}
\end{figure*}

\section{Data sample and analysis procedure}
\label{sec:analysis-procedure}
During the 5-week-long run in 2003, 230 million muon pair ({\it
  dimuon}) triggers were recorded, with two different settings of the
ACM magnet current (4000~A and 6500~A).  The results presented in this
paper refer to the 4000 A data sample, which amounts to $\sim50\%$ of
the total statistics, as the 6500~A setting suppresses the low mass
acceptance and it is optimal for the study of the $J/\psi$.

The data were collected with a $158~\mathrm{A\,GeV}$ indium beam,
incident on 7 indium targets of 17$\%$ total interaction probability.
The targets were placed in vacuum in order not to pollute the data
sample with In-air interactions. The $z$ coordinate of the interaction
vertex is shown in Fig.~\ref{fig:Zvertex}. The resolution on the
position of the interaction vertex is $10 - 20~\mu\mathrm{m}$ in the
transverse coordinates and $\sim 200~\mu\mathrm{m}$ in the
longitudinal coordinate.  The target identification is extremely good,
with negligible background between different targets. Other elements
present along the beam line are also clearly visible in the picture,
like the target box windows and the beam tracker stations.

In the following subsections we illustrate briefly the most relevant
aspects of the reconstruction and selection of events, background
subraction and acceptance correction.

\subsection{Event reconstruction and selection}

As a first step, the muon tracks are reconstructed in the muon
spectrometer.  About half of the triggers are discarded either because
it was not possible to reconstruct two muons or because at least one
of the muons traversed the 4-m-long iron poles of the ACM. These
tracks suffer from a very strong multiple scattering, which would
severly degrade the momentum resolution.  The muon tracks
reconstructed in the muon spectrometer are then extrapolated back to
the target region and mat\-ched to the tracks reconstructed in the
vertex spectrometer. This is done comparing both their angles and
momenta, requiring a {\it matching} $\chi^2$ less than 3. Once
identified, the muons are refitted using the joint information of the
muon and the vertex spectrometers.  These tracks will be referred to
as \emph{matched muons}.  Muon pairs of opposite charge are then
selected.  The matching technique drastically improves the
signal-to-backround ratio and the dimuon mass resolution, which
decreases from $\sim80$~MeV (using only the information from the muon
spectrometer) to $\sim20$~MeV at the \f~peak, independent of
centrality~\cite{Arnaldi:2008er}.

Events with more than one interaction vertex or with a (reconstructed)
charged track multiplicity smaller than 4 are
discarded~\cite{Arnaldi:2008er}.  The matched dimuon mass spectrum is
shown in the left panel of Fig.~\ref{fig:clean_mass}. The same picture
shows also the two background sources: combinatorial and fake matches.
These are discussed in detail in the next subsection.  As it can be
seen, the dominant background is the combinatorial background -- the
fake match background is almost one order of magnitude smaller.  The
$\omega$ and \f~peaks are very well resolved and the $\eta$ peak is
also visible.  The right panel of Fig.~\ref{fig:clean_mass} shows an
expanded view of the $\omega$ and $\phi$ region after the subtraction
of backgrounds. The sample consists of $\sim$70~000 $\phi$ events. The
signal/background ratio, integrated over centrality, is $\sim 1/3$
below the \f~peak, ranging from about 3 in peripheral events to about
1/5 in central ones.

To study $\phi$ production, events in a small mass window centred at
the \f~pole mass ($0.98 < M < 1.06~\gevcs$) are selected from
the mass distribution after the combinatorial background subtraction.
The continuum below the \f~is accounted subtracting the events in two
side mass windows ($0.88 < M < 0.92~\gevcs$ and $1.12 <
M < 1.16~\gevcs$), shown in the right panel of
Fig.~\ref{fig:clean_mass}.

No specific cuts in transverse momentum, rapidity or Collins-Soper
angle are applied. As already mentioned, the detector introduces an
intrinsic rapidity cut $3\lesssim y \lesssim 4.2$ at the $\phi$
mass.

\subsection{Background subtraction}
 
The dimuon sample is affected by two sources of background: the
combinatorial background and the {\it fake} mat\-ches.  The former is
the contribution of uncorrelated muon pairs, mostly coming from the
decay of pions and kaons, and is subtracted with an event mixing
technique~\cite{Arnaldi:2008er,Shahoian:2005ys}. The NA60 apparatus
triggers also on like sign pairs ($\mu^{-}\mu^{-}$ and
$\mu^{+}\mu^{+}$), which, at SPS energies, contain only uncorrelated
muons.  Two muons from different like-sign events are randomly paired
to build a dimuon sample which is uncorrelated by construction.  The
comparison of the real and mixed spectra for the like-sign pairs leads
to an estimate of the accuracy of the background subtraction of $\sim
1\%$ over the full di\-mu\-on mass spectrum.  The like sign sample is
also used to determine the absolute normalization of the combinatorial
background~\cite{Arnaldi:2008er,Shahoian:2005ys}. The combinatorial
background is shown as a dashed line in the left panel of
Fig.~\ref{fig:clean_mass}.

The background from fake track matches comes from the fact that, when
the multiplicity is high, a muon track can be associated to more than
one track in the vertex telescope with an acceptable matching
$\chi^2$, and the track with the lowest $\chi^2$ may not be the
correct match.  It can also happen that the correct track in the
vertex telescope in absent altogether.  The contribution of the fake
matches can be estimated in two different ways.  The first approach is
an overlay Monte Carlo technique, where a Monte Carlo dimuon is
reconstructed on top of a real event. Selecting the match with the
smallest matching $\chi^2$ it is possible to determine the fraction of
fake matches as a function of mass, transverse momentum, etc.  The
left panel of fig.~\ref{fig:MatchingChi2} shows the matching $\chi^2$
for the correctly matched tracks and for the fake matches, as obtained
with an overlay Monte Carlo simulation. The stricter the matching
$\chi^2$ cut, the higher the signal to background ratio will be.
However, a tight cut reduces significantly the statistics. The second
method is based on the event mixing technique, which extracts the
probability distributions of fake matches from real data alone. The
basic idea is to match the tracks in the muon spectrometer from one
event to the vertex tracks of a different
event~\cite{Arnaldi:2008er,Shahoian:2005ys}.  All the matches obtained
in this way are fake by construction. The two methods agree within
5\%.

The right panel of Fig.~\ref{fig:MatchingChi2} shows the Monte Carlo
dimuon invariant mass for \f~events when one or both muons are not
correctly matched, for signal events and for all reconstructed pairs.
This gives an idea how incorrect mat\-ch\-ing degrades the kinematics,
resulting in a broadened \f~pe\-ak with $\sigma\sim100$~MeV. Notice
however, that this is not the total fake matches background under the
\f~peak, because also other processes can contribute - continuum from
open charm or tails from the $\rho-\omega$ region.  The total fake
matches distribution is shown in detail as a dashed line in the right
panel of Fig.~\ref{fig:clean_mass}.  This shows that the fake matches,
in the specific case of the \f~meson, can also be directly eliminated
by the subtraction of the side windows, without the need to estimate
them with the methods outlined above.

In the following, the analysis is repeated subtracting the fakes both
with the event mixing technique and with the side windows.

\subsection{Centrality selection}
\label{sec:CentrSelection}

All results are studied as a function of the collision centrality.
Events are divided into five centrality bins by using the charged
tracks multiplicity as summarised in Table~\ref{tab:inin_centr_bins}.
The reconstructed minimum bias multiplicity distribution depends on
the target because of acceptance, reconstruction efficiency and
secondary particles contamination.  The distributions in the seven
targets are rescaled so that they overlap to that in the first
target. The latter is then fitted with the multiplicity distribution
obtained with a Glauber Monte Carlo simulation, assuming that the
number of reconstructed tracks is proportional to the number of
participants, with a smearing due to the resolution and inefficiencies
of the detector. It is assumed that the effects due to acceptance,
reconstruction efficiency and secondaries are embedded in the
proportionality constant and smearing factor. The distributions and
mean values of the number of participants $N_{part}$ and of the number
of binary collisions $N_{coll}$, are obtained applying the same
multiplicity cuts used in the real data to the simulated data.
Alternatively, the real data can be corrected for acceptance,
reconstruction efficiency and secondaries using a Monte Carlo
simulation~\cite{Floris:Phd,Floris:2005ab}. The number of measured
tracks can be thus converted to the number of primary particles event
by event, as a function of target and centrality. The distribution of
primaries is then matched to the distribution of $N_{part}$ obtained
by a Glauber Simulation. We find $N_{part} \simeq \left. dN_{ch} /
d\eta \right|_{3.7}$~\cite{David:2005ab,Arnaldi:2008er}.

The difference on the mean $N_{part}$ and $N_{coll}$ values for each
bin obtained with the 2 methods is taken as an estimate of the
systematic uncertainty and is found to be about 10\% for the peripheral
bins and 5\% for the central bins.

\begin{table}[ptb]
  \caption{Centrality bins used in the analysis, selected using the charged 
    tracks multiplicity. The intervals used in the five centrality 
    bins are reported in the table, together with the mean values of 
    the number of participants ($N_{\mathrm{part}}$) and of the number of 
    binary collisions ($N_{\mathrm{coll}}$) (see text for details).}
  \label{tab:inin_centr_bins}
  \centering
  \begin{tabular}{ccc}    
    \hline\noalign{\smallskip}
     $dN_{ch}/d\eta$ range  & $\left< N_{\mathrm{part}} \right>$ & $\left< N_{\mathrm{coll}} \right>$ \\
    \noalign{\smallskip}\hline\noalign{\smallskip}
     $  4 - 25 $  &  15   &  14 \\ 
     $ 25 - 50 $  &  41   &  49 \\
     $ 50 - 95 $  &  78   &  110 \\
     $ 95 - 160$  &  133  &  218 \\
     $160 - 250$  &  177  &  317 \\
    \noalign{\smallskip}\hline
  \end{tabular}
\end{table}

\subsection{Acceptance and reconstruction efficiency correction}
\label{sec:Acceptance}
After subtraction of the combinatorial background, the physical
continuum and the fake tracks, the data are corrected for acceptance
and reconstruction efficiency.
\begin{figure}[t]
  \centering
  \includegraphics[width=0.48\textwidth,height=0.48\textwidth]{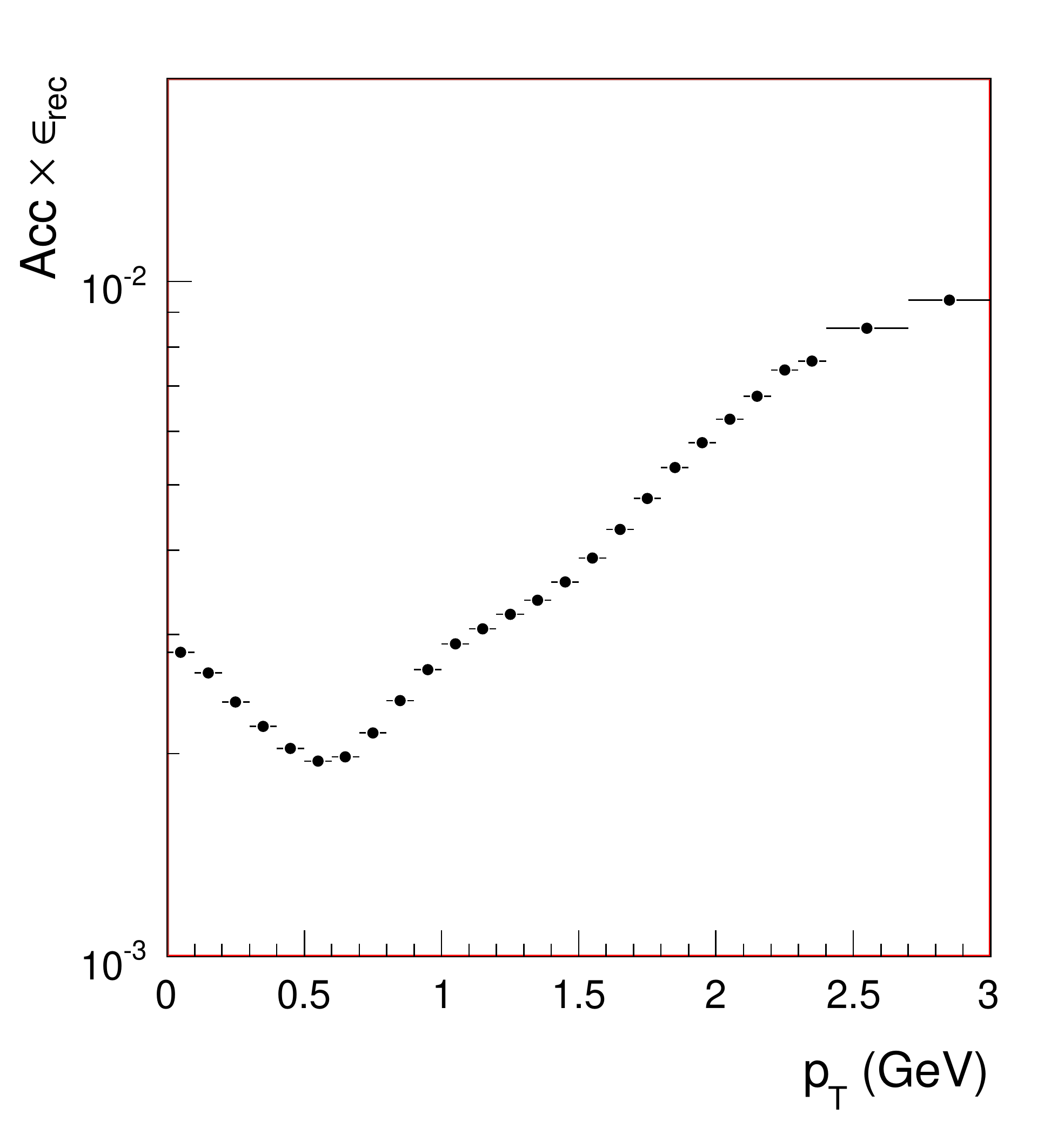}
  \caption{Geometrical acceptance times reconstruction efficiency
    as a function of \pt~for the \f.}
  \label{fig:acceptance}
\end{figure}
These are estimated with an overlay Monte Carlo simulation, using the
real hit occupancy in the detector superimposed to Monte Carlo muon
pairs.  The same cuts used for the real data were applied to the
simulated data. An overall 95\% efficiency was assumed for the pixel
detector.  The input sources' kinematics are generated using the
Genesis event generator~\cite{genesis}, tracked through the apparatus
using GEANT 3.21 and re\-con\-struc\-ted using the NA60 analysis
framework.

The study of the differential spectra would require a multi
dimensional acceptance correction in rapidity, transverse momentum and
Collins-Soper angle.  Since the low populated bins in the phase space
corners can introduce large errors, a fiducial cut should be applied,
at the cost of reducing the effective statistics.  To avoid this loss,
a one dimensional correction is applied to each kinematic variable
under study, integrating over the remaining ones. This requires a
careful tuning of the Monte Carlo input kinematics which was adjusted
with an iterative procedure to the measured distributions for each of
the five centrality bins.  To validate this procedure, the full
analysis chain (event selection, reconstruction, acceptance
correction, fitting) is applied to Monte Carlo events. It is verified
that the corrected kinematic distributions and the estimated
parameters are consistent with those used as input in the Monte Carlo.

The acceptance times reconstruction efficiency for the $\phi$ as a
function of $p_T$ is shown in Fig.~\ref{fig:acceptance}. The apparatus
has a good acceptance down to zero \pt, due to the additional dipole
field in the vertex region. Indeed, low mass and low \pt~dimuons,
which would not be accepted by the muon spectrometer, are bent into
the acceptance by the vertex dipole magnet, as opposed to NA50, which
used the same muon spectrometer, but had no acceptance for
$p_T\lesssim 1.1~\mathrm{GeV}$. Further details on the rapidity and
angular distribution acceptance will be given in the specific
subsections on the results.

\subsection{Systematic effects}
\label{sec:systematic-effects}

The main sources of systematic error are the uncertainties on the
input parameters of the Monte Carlo simulation (which reflect on an
uncertainty in the acceptance) and on the estimate of the physical
background below the $\phi$. The former is estimated by changing the
values of the parameters in the input Monte Carlo within $1\sigma$ of
the measured value. The latter, by varying the side windows offset
(from 50~MeV to 70~MeV) and width (from 30 MeV to 50 MeV). The
contributions from these two factors are added in quadrature.
Uncertainties in the combinatorial background and in the fake matches
subtraction play a small role here, as the $\phi$ peak is very well
resolved.
\label{sec:rapid-distr}
\begin{figure}[t]
  \centering
  \includegraphics[width=0.48\textwidth,height=0.48\textwidth]{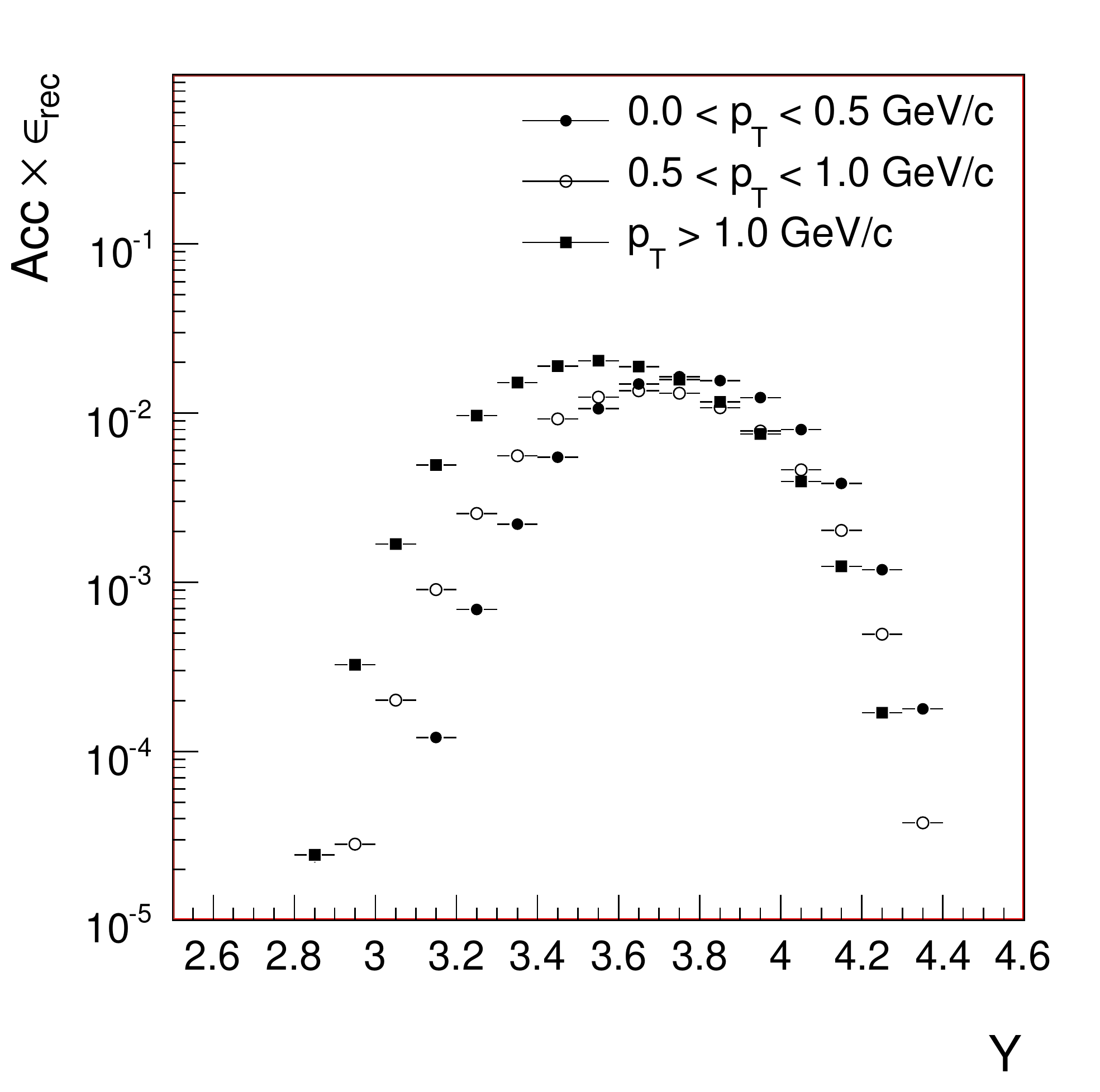}
  \caption{Geometrical acceptance times reconstruction efficiency as a
    function of y for the \f~for $0<p_T<0.5$, $0.5<p_T<1$ and $p_T>1$
    GeV.}
  \label{fig:accy}
\end{figure}

Several consistency checks were performed to test for the presence of
hidden systematic effects~\cite{Barlow}: the matching $\chi^2$ cut was
varied between 2 and 3, the kinematic cuts $ 2.9 < y < 3.9$ and
$\left| \cos\theta_{CS}\right| < 0.5$ were applied, the analysis was
repeated independently in the 4 ACM/PT7 fields' combinations, a
2-dimensional acceptance correction (vs. $y$ and $p_T$) was used and
the different methods for fake subtraction were implemented. Most of
those checks did not lead to significantly different results, with the
exception of the matching $\chi^2$ cut variation. This leads to a
small but not negligible difference which is added in quadrature to
the systematic error. This difference is most likely due to the
uncertainty in the energy loss compensation or to a small residual
misalignment in the pixel telescope planes.
\begin{figure*}[t]
  \centering
  \subfigure{\includegraphics[width=0.48\textwidth,height=0.48\textwidth]{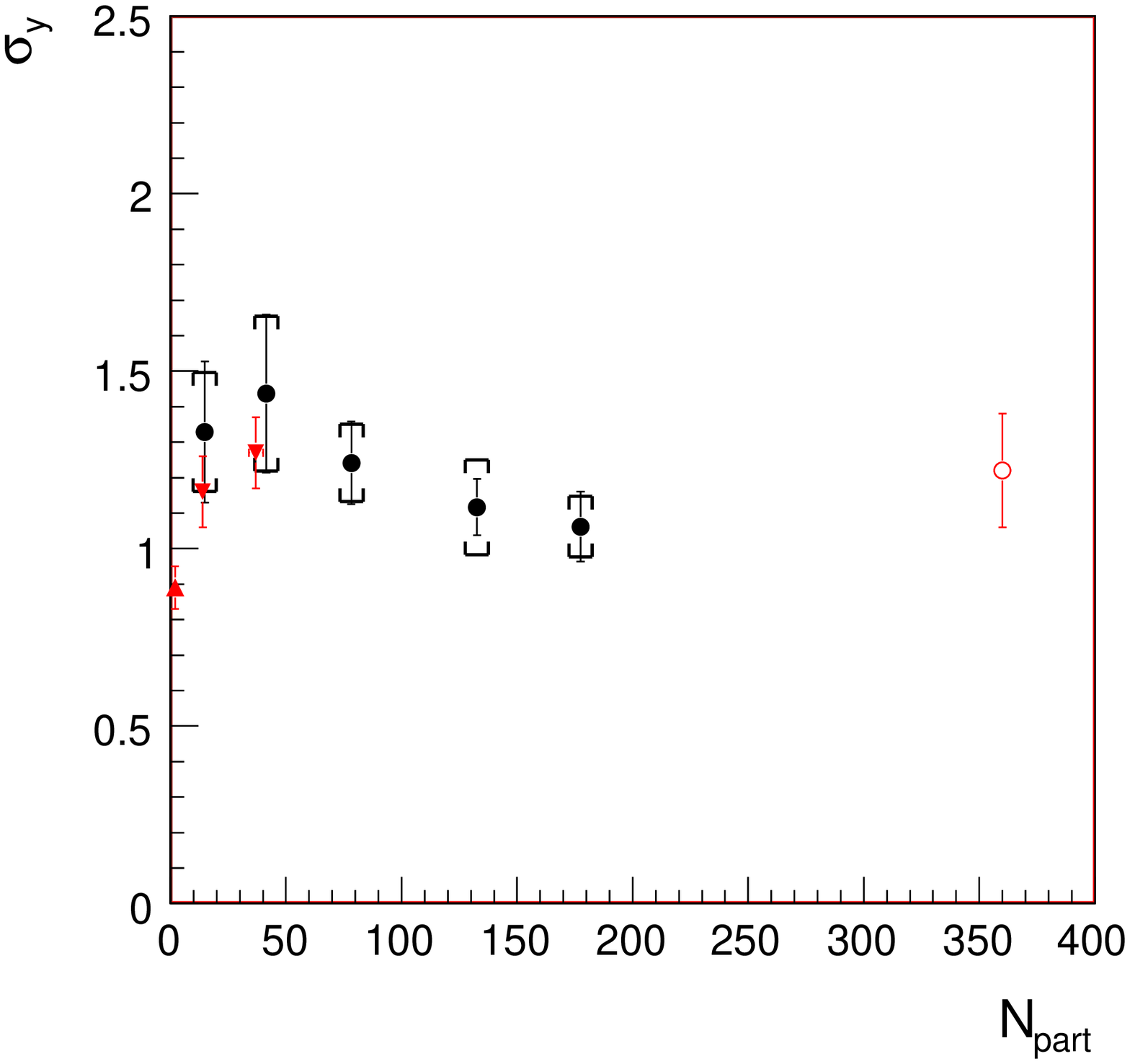}}
  \subfigure{\includegraphics[width=0.48\textwidth,height=0.48\textwidth]{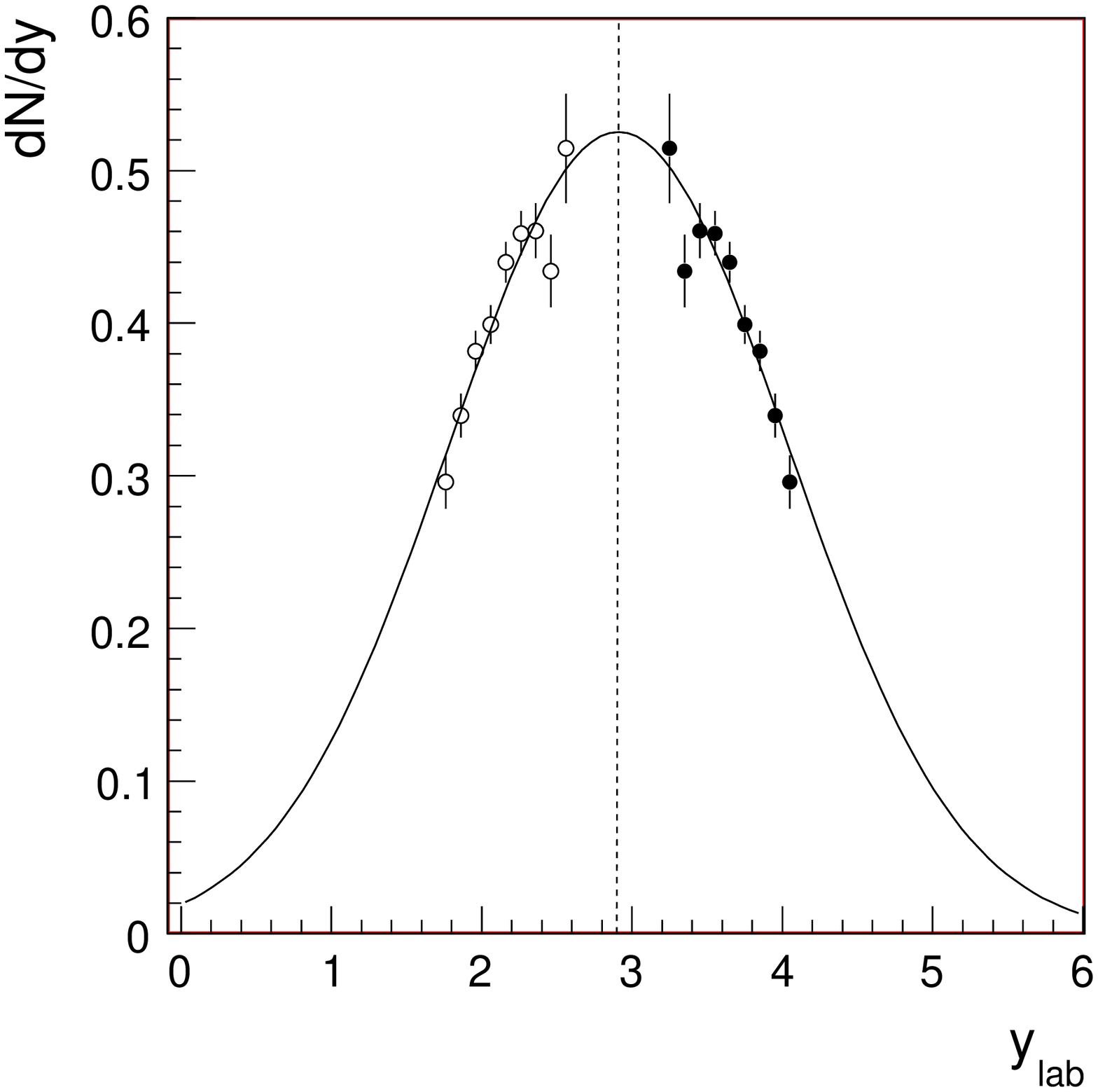}}

  \caption{Left panel: Gaussian $\sigma$ of the rapidity distribution
    of the \f~meson, as a function of centrality; full circle:
    $\phi\to\mu\mu$ in In-In (NA60); full upward triangle: $\phi\to
    K^+K^-$ in p-p (NA49); full downward triangle: $\phi\to K^+K^-$ in
    C-C and Si-Si (NA49); open circle: $\phi\to K^+K^-$ in Pb-Pb
    (NA49). Right Panel: Rapidity distribution of the \f~meson; open
    symbols are obtained reflecting the measured points around
    mid-rapidity.}
  \label{fig:rapidity}
\end{figure*}
The results on the yield, mass and width are affected by further
sources of systematics, due to the parameters needed for their
computation. Details are given in sec.~\ref{sec:yield} and in
sec.~\ref{sec:Mass_Width}.  The number of participants estimated with
the Glauber model is affected by a 10\% uncertainty in the peripheral
bins, which decreases to $\sim5\%$ in the most central, as explained
in sec.~\ref{sec:CentrSelection}.

\section{Results}
\label{sec:differential-spectra}

\subsection{Rapidity}

The rapidity acceptance for $0<p_T<0.5$, $0.5<p_T<1$ and $p_T>1$
GeV is shown in Fig.~\ref{fig:accy}. The apparatus has acceptance
in the interval 3-4.2 depending on the target. There is a small
correlation $y-p_T$ such that low $y$ tends to select larger $p_T$.
The rapidity is studied as a function of centrality and the width
estimated with a gaussian fit in the range $3.2 < y < 4.1$.  The
midrapidity value is fixed to 2.91.  The width as a function of
centrality is shown in Fig.~\ref{fig:rapidity} (left panel) and within
the errors one can deduce either a constant or a slightly decreasing
trend.  The rapidity distribution integrated in centrality is shown in
Fig.~\ref{fig:rapidity} (right panel). The resulting width is $ 1.13
\pm 0.06 \pm 0.09$. This is in agreement with previous NA49
measurements in other collision systems at the same
energy~\cite{Alt:2004wc,Afanasev:2000uu}, as shown in
Table~\ref{tab:sigma-y-previous}.

\begin{table}[ptb]
  \caption{Gaussian width of the rapidity distribution of the \f~meson
    in several collision systems, at 158~AGeV incident beam
    energy. The second quoted error for the NA60 point is the
    systematic error.}
  \label{tab:sigma-y-previous}
  \centering
  \begin{tabular}{c c c}
    \hline\noalign{\smallskip}
    Experiment & System & $\sigma_{y}$  \\
    \noalign{\smallskip}\hline\noalign{\smallskip}
    NA49 & p-p   & $0.89 \pm 0.06$ \\
    NA49 & C-C   & $1.16 \pm 0.10$ \\
    NA49 & Si-Si & $1.27 \pm 0.10$ \\
    NA60 & In-In & $1.13 \pm 0.06 \pm 0.09 $\\
    NA49 & Pb-Pb & $1.22 \pm 0.16$ \\
    \noalign{\smallskip}\hline
    
  \end{tabular}
\end{table}

\subsection{Angular distributions}
\label{sec:angular-distribution}

\begin{figure*}[tbp]
  \centering
  \subfigure{\includegraphics[width=0.48\textwidth,height=0.48\textwidth]{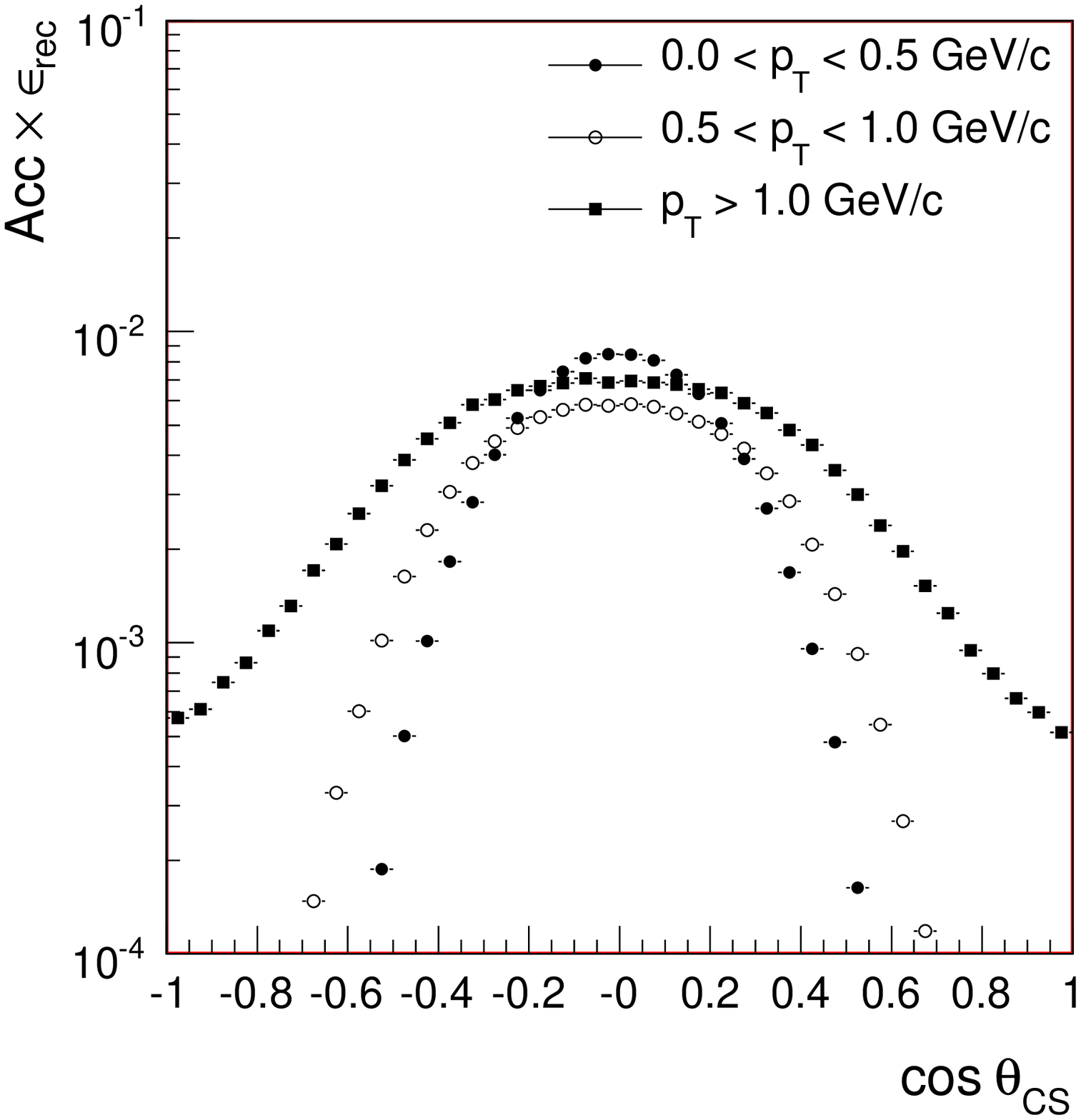}}
  \subfigure{\includegraphics[width=0.48\textwidth,height=0.48\textwidth]{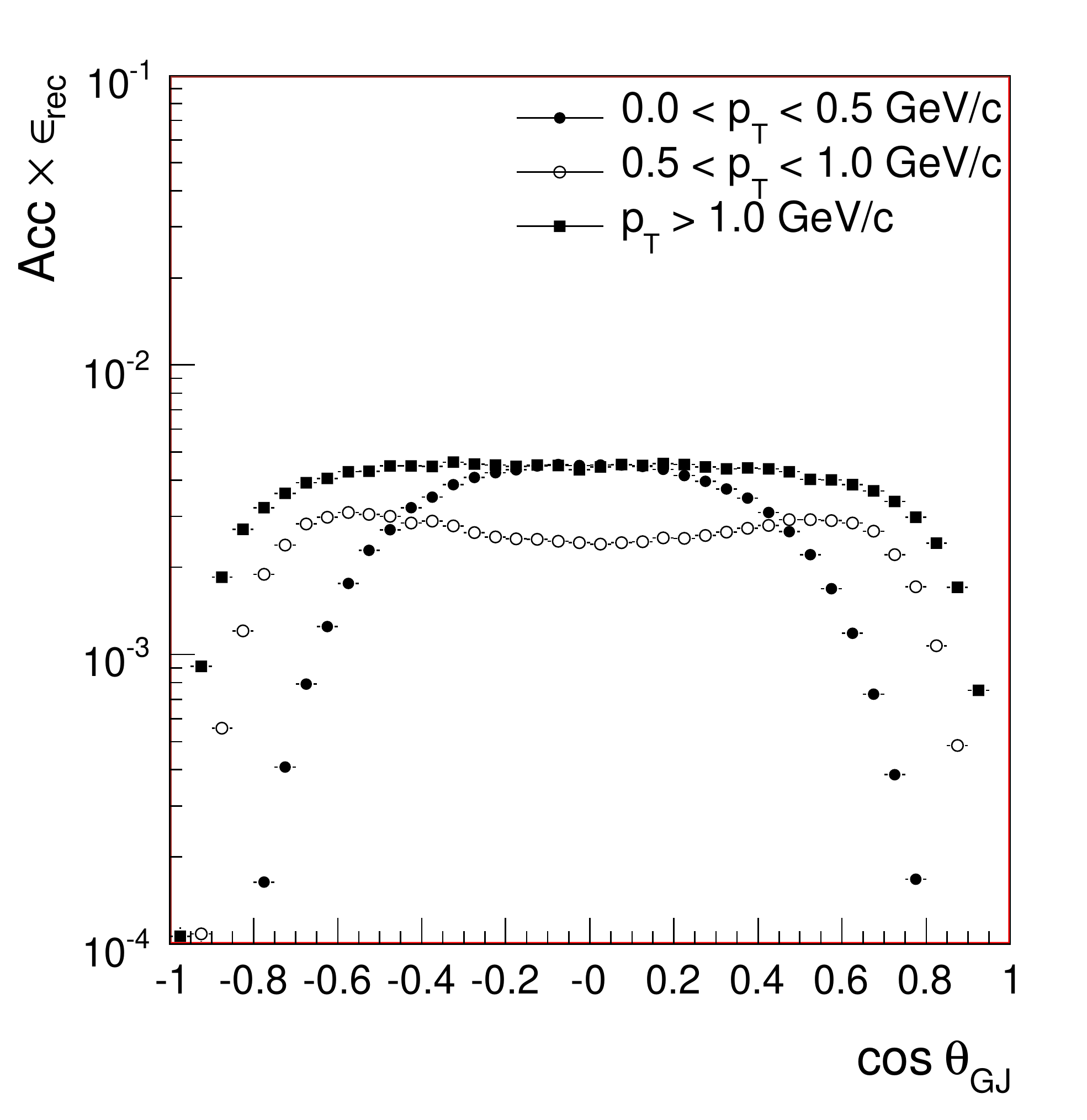}}
  \caption{Geometrical acceptance times reconstruction efficiency as a
    function of $\cos\theta$ for the \f~for $0<p_T<0.5$, $0.5<p_T<1$
    and $p_T>1$ GeV in the CS frame (left) and in the GJ frame
    (right).}
  \label{fig:CSacceptance}
\end{figure*}

\begin{figure*}[pbt]
  \centering
  \subfigure{\includegraphics[width=0.48\textwidth,height=0.48\textwidth]{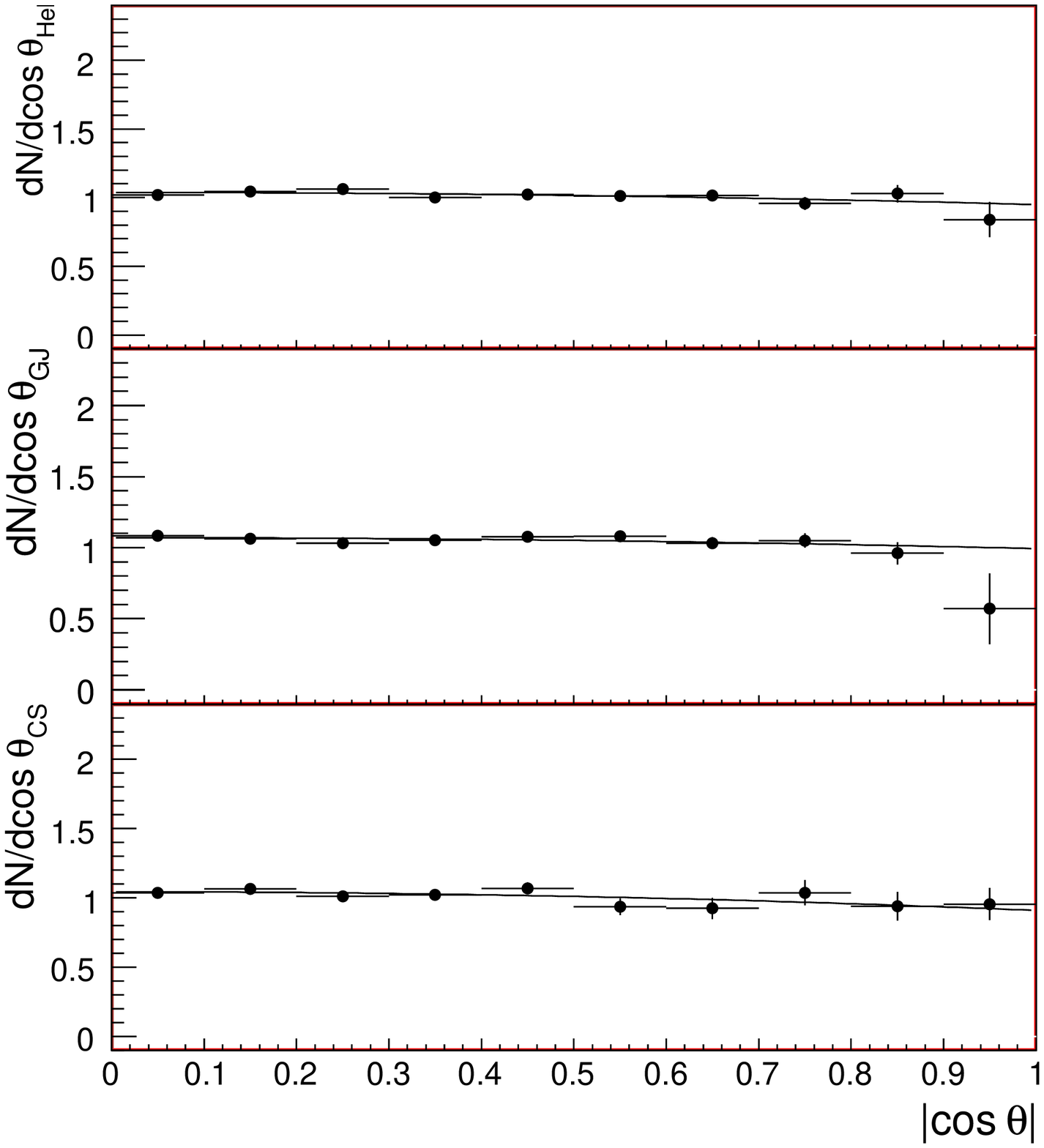}}
  \subfigure{\includegraphics[width=0.48\textwidth,height=0.48\textwidth]{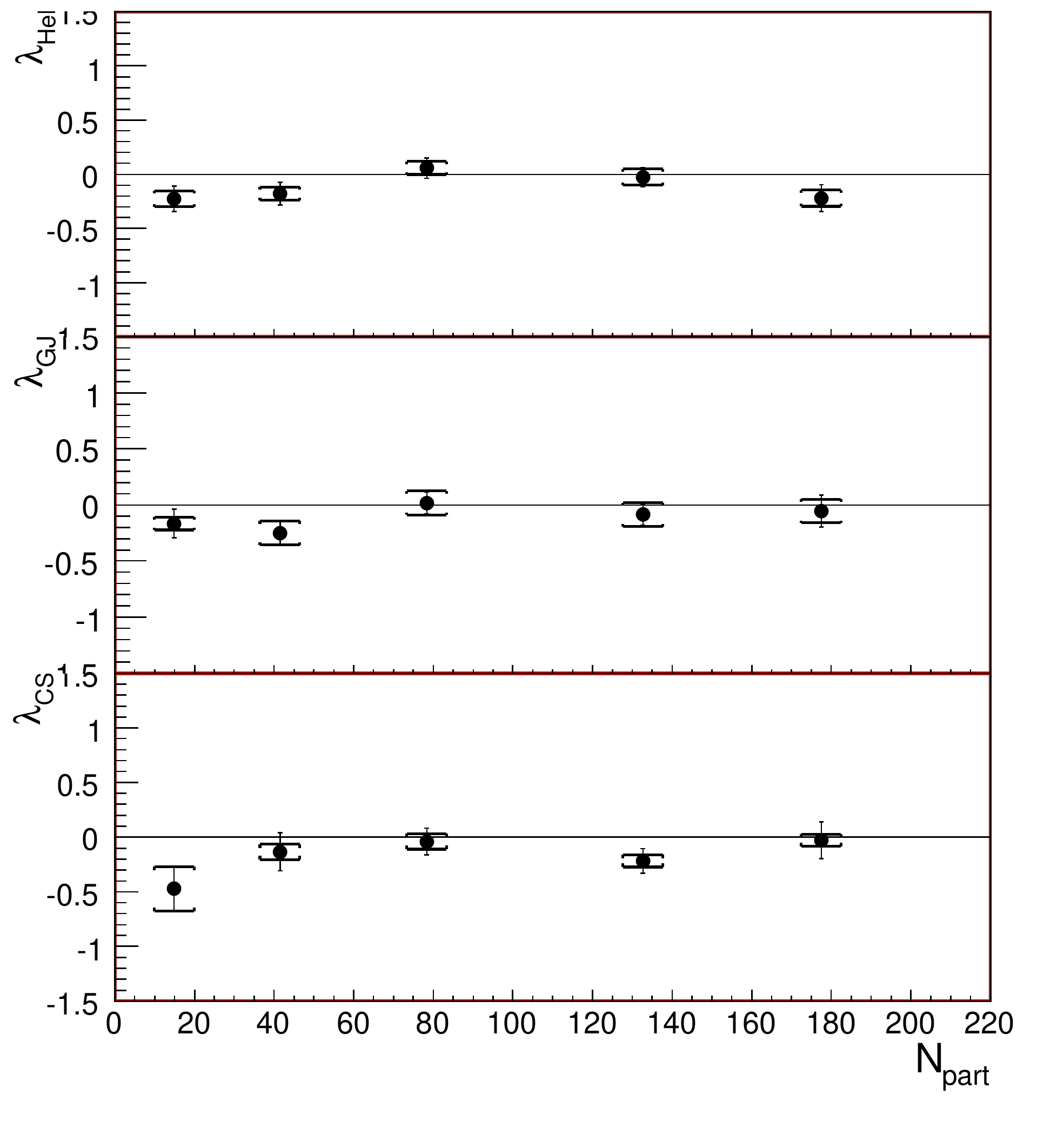}}
  \caption {Angular distributions of the \f, measured in the helicity,
    Gottfried-Jackson and Collins-Soper frames. Left panels: angular
    distributions integrated in centrality. Right panels: polarization
    as a function of centrality.}
   \label{fig:angular}
\end{figure*}

In general, the angular distribution of the decay muons in the $\phi$
meson rest frame can be related to the elements of the spin density
matrix.

In this paper the inclusive $dN/d\cos\theta$ distributions are
studied as a function of transverse momentum and centrality.  The
polar angle $\theta$ is the angle between the $\mu^{+}$ and the $z$
axis as measured in the following three widely used reference frames.
In the helicity frame, the axis is defined as the direction of flight
of the $\phi$ meson in the target-projectile rest frame. In the
Gottfried-Jackson frame (GJ), the axis is defined as the projectile
direction in the \f~rest frame~\cite{Gottfried:1964nx}. Finally, in
the Collins-Soper frame (CS) the axis is given by the bisector of the
projectile and the negative target direction in the \f~rest
frame~\cite{Collins:1977iv}.

In all the three frames, the \f~acceptance as a function of
$\cos\theta$ has a strong dependence on transverse momentum. In
particular, the apparatus has zero acceptance for high
$\left|\cos\theta\right|$ and low $p_T$.  The acceptance vs
$\cos\theta$ for the \pt~windows $0<p_T<0.5$, $0.5<p_T<1$ and $p_T>1$
GeV, is shown in Fig.~\ref{fig:CSacceptance} for the CS and GJ
frames. The most critical situation occurs for the Collins-Soper
angle, while the helicity frame is similar to the GJ frame.

The $dN/d\cos\theta$ distribution is fitted with the function $1+
\lambda\cdot\cos^2\theta,$ where the parameter $\lambda$ is related to
the degree of polarization.  The spectra, integrated in centrality and
$p_T$, are shown in the left panel of Fig.~\ref{fig:angular}. The
polarizations obtained from the fits are zero within errors in all
cases: $\lambda_{HEL} = -0.07 \pm 0.06 \pm 0.06$, $\lambda_{GJ} =
-0.07 \pm 0.06 \pm 0.06$, $\lambda_{CS} = -0.13 \pm 0.07 \pm 0.06$.

The polarization is then studied differentially in the three $p_T$
windows $0<p_T<0.5$, $0.5<p_T<1$ and $p_T>1$ GeV, without
centrality selection.  No significant deviations from the $p_T$
integrated values are found.

The results as a function of centrality and integrated in $p_T$ are
shown in the right panels of Fig.~\ref{fig:angular}. The polarization
is compatible with zero in the helicity and Gottfried-Jackson frames,
independent of centrality.  In the Collins-Soper frame there is a hint
for a negative polarization in the most peripheral bin.  However, this
is within $2\sigma$ compatible with zero and then consistent with the
absence of any effect as seen in the other frames.

These results as a function of $p_T$ and centrality are consistent
with the results integrated in centrality with
$p_T>0.6~\mathrm{GeV}$ reported in a previous analysis by
NA60~\cite{Arnaldi:2008gp}. They show the absence of polarization,
supporting the fact that particles are produced from a thermalized
medium. The absence of polarization extends down to peripheral events,
suggesting that it could be zero also in elementary collisions at this
energy.

\subsection{Transverse momentum}
\label{sec:transv-moment-distr}

\begin{figure*}[ptb]
  \centering
  \subfigure{\includegraphics[width=0.48\textwidth,height=0.48\textwidth]{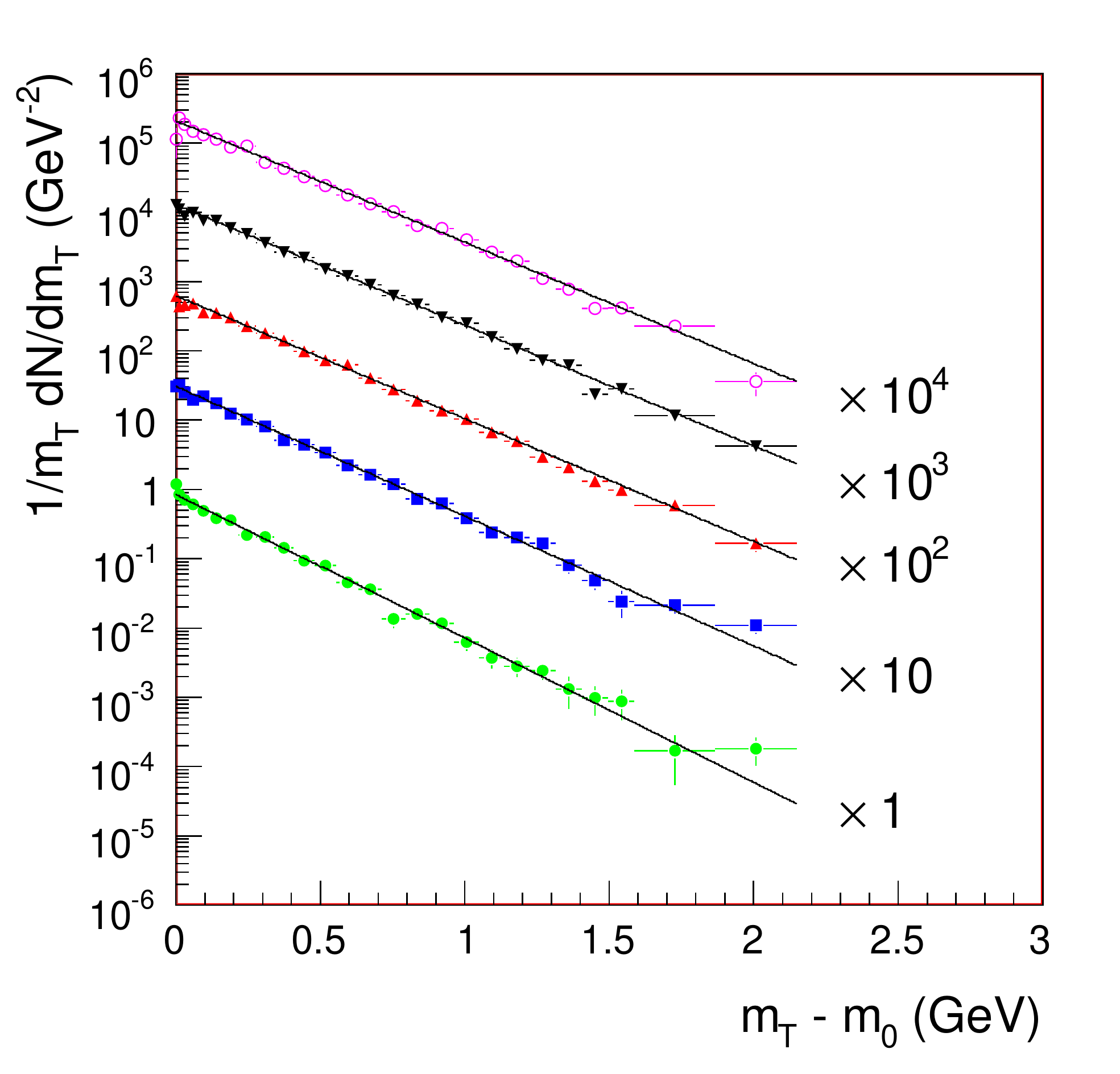}}
  \subfigure{\includegraphics[width=0.48\textwidth,height=0.48\textwidth]{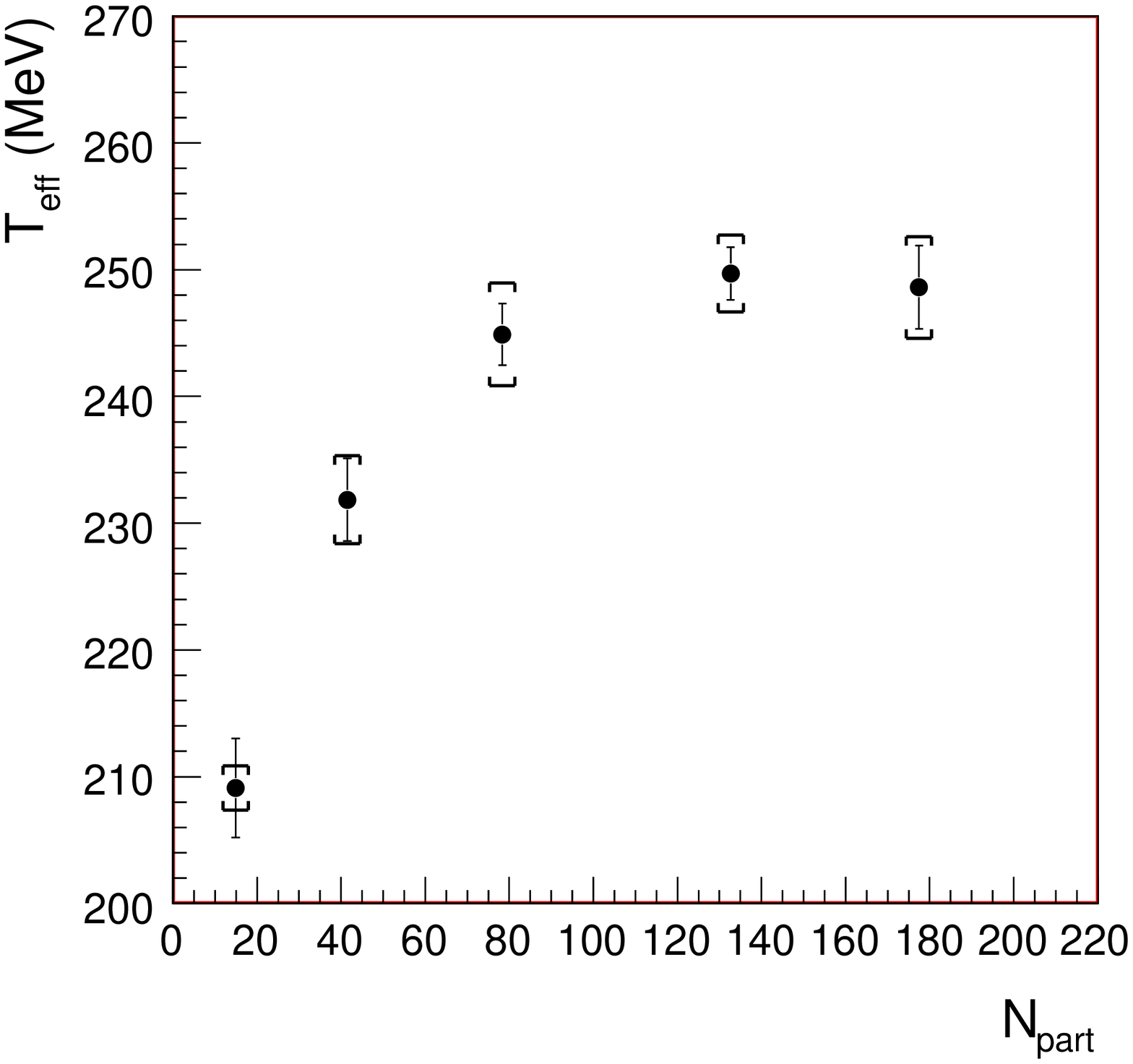}}

  \caption{Left panel: \f~transverse mass distributions in
    indium-indium collisions as a function of centrality; from top to
    bottom: central to peripheral spectra. Right Panel: Centrality
    dependence of the $T_{eff}$ parameter; fit performed over the full
    NA60 range ($0 < p_\mathrm{T} < 3~\gevc$).}
  \label{fig:mt-spectra-and-tfull}
\end{figure*}

\begin{table*}[ptb]
  \caption{Centrality dependence of the $T_{eff}$ parameter in
  different fit ranges. The second quoted error is the systematic
  error.}
  \label{tab:tslope-fit-ranges}
  \centering
  \begin{tabular}{c c c | c c }
    \hline\noalign{\smallskip}
    \multirow{2}{*}{$\left< N_{\mathrm{part}} \right>$} &  $T_{eff}$~(MeV) & $T_{eff}$~(MeV) &  
    \multirow{2}{*}{$\left< N_{\mathrm{part}} \right>$} & $T_{eff}$~(MeV) \\
    &  $0 < p_T < 3$~GeV & $0 < p_T < 1.6$~GeV & & $1.1 < p_T < 3$~GeV \\
    \noalign{\smallskip}\hline\noalign{\smallskip}
    15  &$ 209 \pm 4 \pm 2 $  & $ 206 \pm 5 \pm 2 $ & \multirow{2}{*}{24}  &    \multirow{2}{*}{$ 228 \pm 6 \pm 4 $}\\
    41  &$ 232 \pm 3 \pm 3 $  & $ 229 \pm 5 \pm 2 $ &                                                            &  \\
    78  &$ 245 \pm 2 \pm 4 $  & $ 253 \pm 5 \pm 9 $ & 78  &                                     $ 237 \pm 5 \pm 1 $ \\
   133  &$ 250 \pm 2 \pm 3 $  & $ 258 \pm 5 \pm 5 $ & 133 &                                     $ 245 \pm 4 \pm 5 $ \\
   177  &$ 249 \pm 3 \pm 4 $  & $ 253 \pm 8 \pm 11$ & 177 &                                     $ 244 \pm 5 \pm 3 $ \\
    \noalign{\smallskip}\hline
  \end{tabular}
\end{table*}

The acceptance-and-efficiency corrected transverse mass distributions
are shown in the left panel of Fig.~\ref{fig:mt-spectra-and-tfull} for
the different collision centralities, normalized in absolute terms.

The transverse momenta are first fitted in the range $0 < p_T <
3~\mathrm{GeV}$ with the thermal ansatz:
\begin{equation}
\frac{1}{m_T}~\frac{dN}{dm_T} = \frac{1}{p_T}~\frac{dN}{dp_T} \propto \exp(-m_T/T_{eff}).
\label{eq:expo}
\end{equation}
The effective temperature $T_{eff}$ as a function of the number of
participants is depicted in the right panel
Fig.~\ref{fig:mt-spectra-and-tfull}. In general one can notice that
the $T_{eff}$ shows an initial rise followed by a saturation in the
most central bins.

The exponential form assumes thermal emission from a static source. In
heavy ion collisions, this is not correct due to the presence of
radial flow, but eq.~(\ref{eq:expo}) was nevertheless used by several
experiments to fit the data. Since the $m_T$ distributions are not
purely exponential, The $T_{eff}$ extracted from an exponential fit
will depend in general on the fitted $p_T$ range.  It can be shown
that, at first order, the $T_{eff}$ value at low $p_T$ is given by
$T_{eff}\sim T_0+m\beta^2_T$, where $T_0$ is the temperature at
freeze-out, $m$ is the particle mass and $\beta_T$ is the average
transverse velocity of the expanding source~\cite{Heinz:2004qz}. At
high $p_T$, on the other hand, the $T_{eff}$ extracted from the
exponential fit is approximated by $T\sim
T_0\sqrt{(1+\beta_T)/(1-\beta_T)}$~\cite{Heinz:2004qz}.  Thus, what is
measured is a $p_T$ dependent effective temperature, larger than the
freeze-out temperature because of the radial flow.

The influence of radial flow is investigated by repeating the fits in
two sub-ranges $0 < p_T < 1.6~\mathrm{GeV}$ (low $p_T$) and $1.1 <
p_T < 3~\mathrm{GeV}$ (high $p_T$), which also correspond to the
experimental windows covered by the NA49 and NA50 experiments,
respectively. Since the statistics in the peripheral data at high
\pt~is limited, the two most peripheral bins are integrated for the
fit in the high $p_T$ range.  The numerical values of $T_{eff}$ as a
function of centrality and of the different fit ranges are summarised
in Tab.~\ref{tab:tslope-fit-ranges}. An increase of the average
$T_{eff}$ can be noticed when the fit is restricted to the low $p_T$
range, while a flatter trend is observed from the fit in the high
$p_T$ range.  The difference between the effective temperatures
obtained in the two fit ranges is around 10 MeV (average over the
three most central bins). This pattern indicates the presence of
radial flow but it is rather small, as expected from the small
coupling of the \f~to the ``pion
wind''\cite{Teaney:2001av,Damjanovic:2008ta}.

\begin{figure}[tbp]
  \centering
  \includegraphics[width=0.48\textwidth,height=0.48\textwidth]{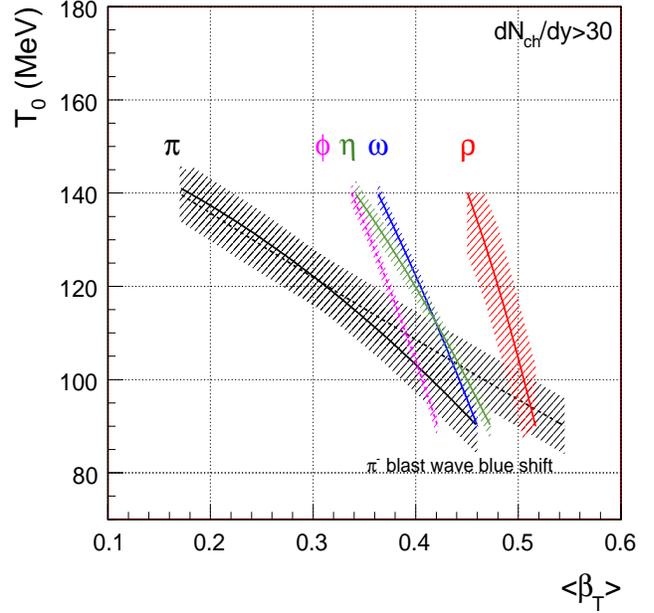}  
  \caption{Blast wave analysis results for $\eta$, $\rho$, $\omega$,
    $\phi$ and negative hadrons (pions). Fits performed in the range
    $0.4 < p_{T} < 1.8~\mathrm{GeV}$.}
  \label{fig:blast_wave}
\end{figure}

Further quantitative insight has been gained studying the spectra in
the framework of the blast wave model~\cite{Schnedermann:1993ws}:
\begin{equation}
\frac{dN}{m_{T}dm_{T}} \propto \int_0^R r dr m_{T} I_0 \left( \frac{p_{T}\sinh \rho}{T_0} \right) K_1 \left( \frac{m_{T}\cosh \rho}{T_0} \right),
\label{eq:BW}
\end{equation}
where $\rho = \tanh^{-1} \beta_r(r)$ and a linear velocity profile was
assumed: $\beta_r = \beta_s \left(r/R\right)$.  The slopes of the
$p_{T}$ spectra define lines in the $T_{0}-\beta_T$ plane. $T_{0}$ is
varied in the window 90-140~MeV. Transverse momentum distributions are
then generated using eq.~(\ref{eq:BW}), with $\beta_T$ values tuned in
order to reproduce the observed $T_{eff}$ values.
Fig.~\ref{fig:blast_wave} shows the results for the $\phi$ and also
for the negative charged hadrons (essentially pions), $\eta$, $\omega$
and $\rho$~\cite{Damjanovic:2008ta}. The study is performed for
$p_T>0.4~\mathrm{GeV}$ in order to exclude the rise in the low $p_T$
region seen in the $\rho$ excess spectra~\cite{Arnaldi:2007ru}. In
addition, to perform a comparison with meaningful statistical accuracy
with the other particles, data were integrated in centrality,
excluding the most peripheral events.  A clear hierarchy in the freeze
out parameters, as a consequence of the particle coupling to the
medium is visible. The $\phi$ is the least coupled particle, while the
$\rho$, being continuously produced in the medium via $\pi^+\pi^-$
annihilation is maximally coupled. The freeze-out component of the
$\rho$ receives the strongest radial flow boost, with a $T_{eff} >
300$~MeV, 50 MeV above the
$\phi$~\cite{Damjanovic:2008ta,Arnaldi:2007ru}.

\begin{figure*}[ptb]
  \centering
  \subfigure{\includegraphics[width=0.48\textwidth,height=0.48\textwidth]{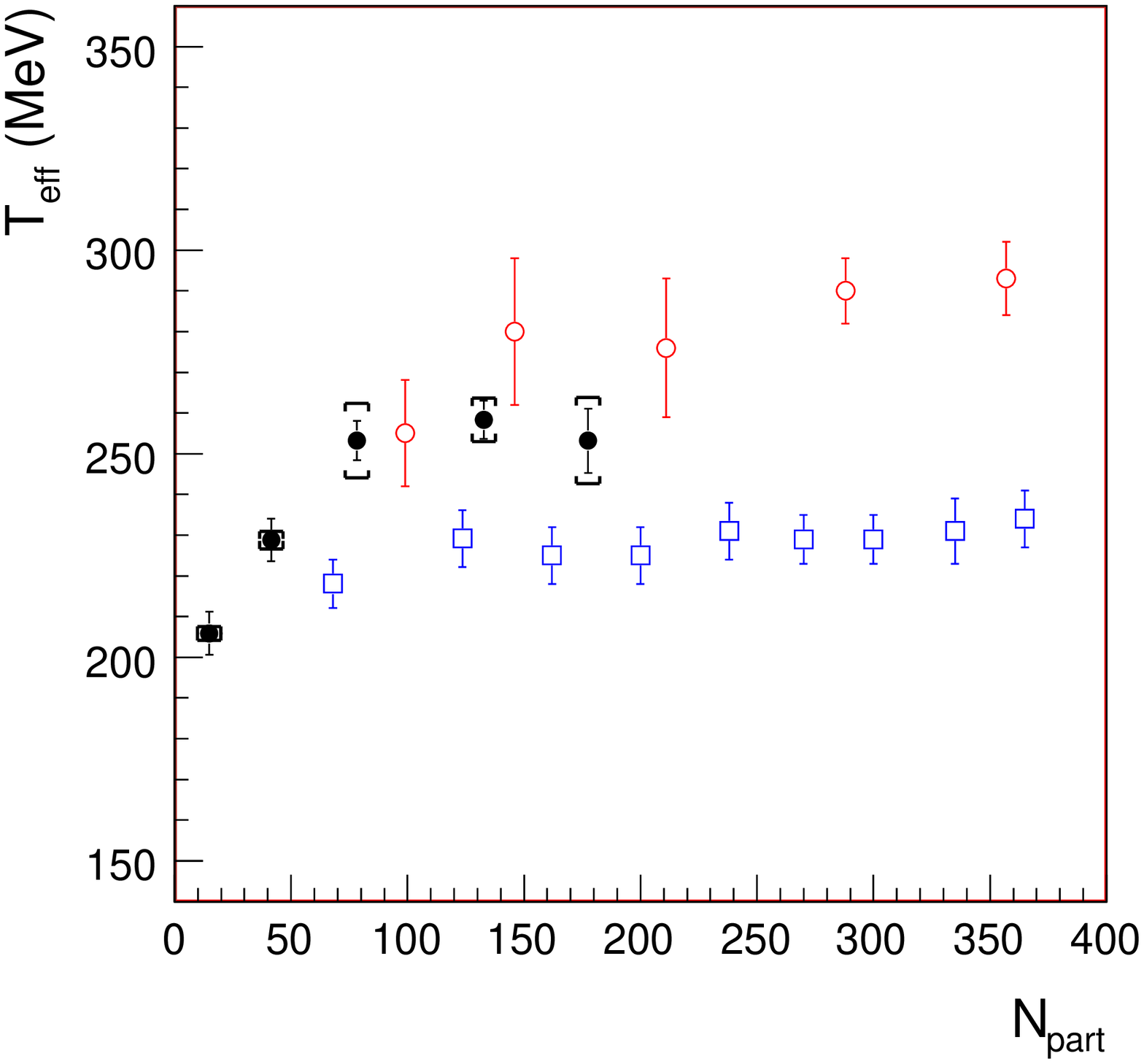}}
  \subfigure{\includegraphics[width=0.48\textwidth,height=0.48\textwidth]{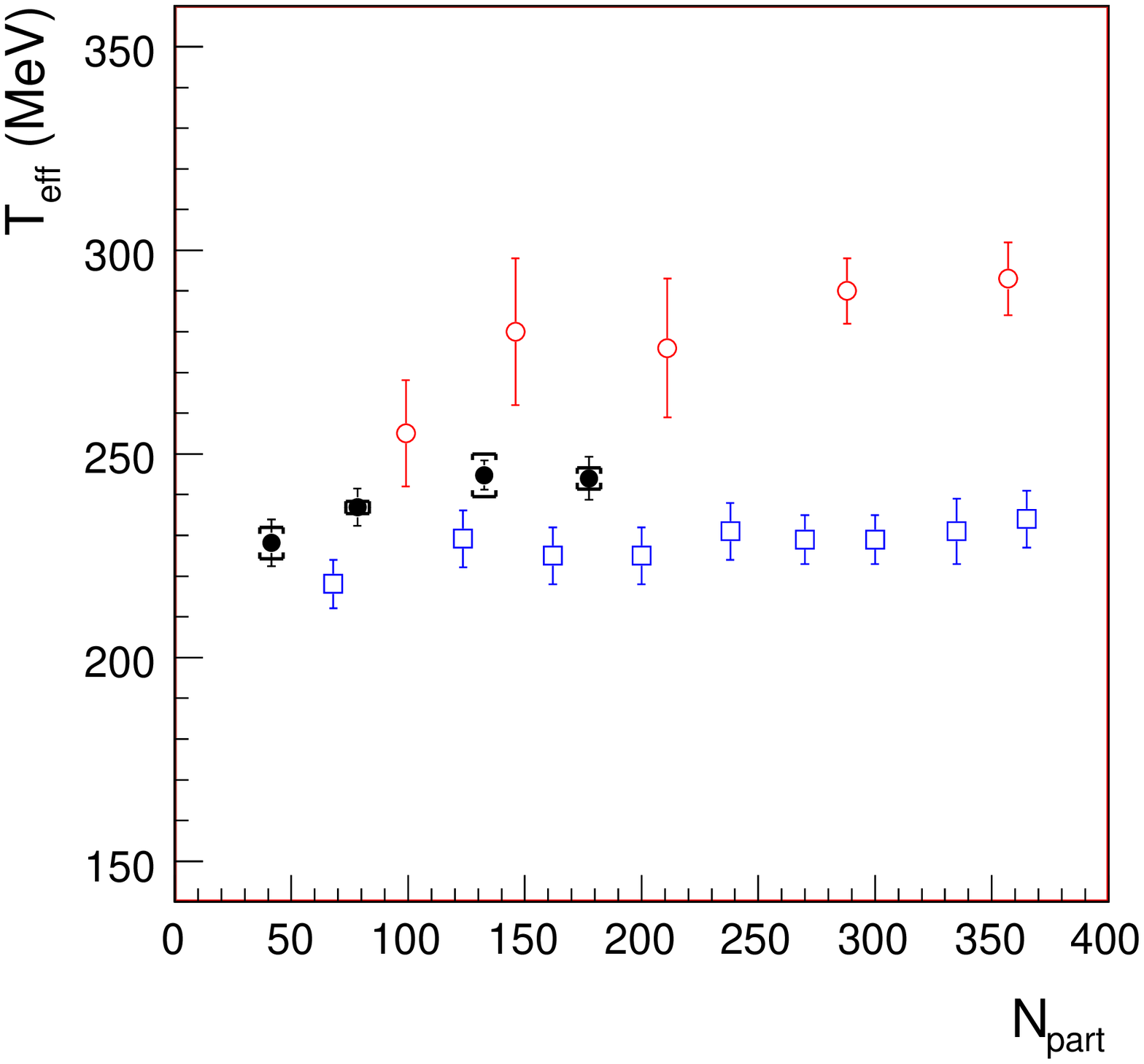}}

  \caption{Centrality dependence of the $T_{eff}$ parameter (full
    circles) compared to NA49 (open circles) and to NA50 (open
    squares). Left panel: fit performed in the NA49 range ($0 <
    p_\mathrm{T} < 1.6~\gevc$).  Right panel: fit performed in the
    NA50 range ($p_\mathrm{T}  > 1.1~\gevc$). }
  \label{fig:T_slope_na49na50range}
\end{figure*}

\subsection{Yield}
\label{sec:yield}

The average number of $\phi$ mesons per interaction
$\left<\phi\right>$ is evaluated using the $J/\psi$ as a reference
process.  Once corrected for anomalous and nuclear absorption,
$\left<J/\psi\right>$ scales with the number of binary collisions and
can be written as 
\begin{equation}
\left<J/\psi\right>_{\mu\mu} =
(\sigma_{NN}^{J/\psi,\mu\mu}/\sigma_{NN})\cdot N_{coll}.
\label{eq:psi-vs-ncoll}
\end{equation}
Here $\sigma_{NN}^{J/\psi,\mu\mu}=5.64\pm0.1$ nb is the $J/\psi$ cross
section in nucleon-nucleon collisions multiplied for the
$J/\psi\to\mu\mu$ branching ratio. It was derived by the systematic
study of $J/\psi$ production in pA collisions performed at 400 and 450
GeV, rescaling to 158 GeV and extrapolating to full phase
space~\cite{goncalo}.  The estimated nucleon-nucleon inelastic cross
section at $\sqrt{s} = 17.2$~GeV is
$\sigma_{NN}=31.7\pm0.5$~mb~\cite{Amsler:2008zz}.

$\langle\phi\rangle$ can be evaluated from the experimental ratio:
\begin{equation}
\label{eq:ratio}
\frac{\left<\phi\right>}{\left<J/\psi\right>_{\mu\mu}} =
\frac{N_{\phi}/\left( \mathrm{A}_{\phi}\times\varepsilon_{rec}^{\phi}\times BR^{\phi}_{\mu\mu} \right)}
{N_{J/\psi}/\left( \mathrm{A}_{J/\psi}\times\varepsilon_{rec}^{J/\psi}\times f^{Abs}_{J/\psi} \right)}
\end{equation}
where $N_{\phi(J/\psi)}$ is the number of $\phi~(J/\psi)$ mesons
observed in a given centrality bin,
$\mathrm{A}_{\phi(J/\psi)}\times\varepsilon_{rec}^{\phi(J/\psi)}$ are
the acceptance and reconstruction efficiency, $BR^{\phi}_{\mu\mu}$ is
the $\phi \to \mu\mu$ branching ratio and $f^{Abs}_{J/\psi}$
incorporates the $J/\psi$ nuclear and anomalous absorption
factors. The same centrality bins, selected using charged particle
multiplicity as described in sec.~\ref{sec:CentrSelection}, were used
for both the $\phi$ and the $J/\psi$. Invoking lepton universality,
the value $2.97\times10^{-4}$ measured in the $\phi \to ee$ is used
for the branching ratio, since it is known with better
precision~\cite{Amsler:2008zz}.  The nuclear absorption was accurately
determined by several measurements in pA collisions at 400 and 450
GeV~\cite{goncalo}. The anomalous absorption was determined from the
$J/\psi$ analysis of the same In-In data sample~\cite{Arnaldi:2007zz}.
The overall correction factor comes from the product of the nuclear
and anomalous absorption. One has to notice that recent preliminary
p-A results from NA60 showed the nuclear absorption at 158~GeV is
larger than previously assumed~\cite{enrico-qm09}. However, since the
anomalous absorption is defined relative to the nuclear one, the
overall correction $f^{Abs}_{J/\psi}$ remains largely insensitive to
relative changes of these two factors. The systematic uncertainty
arising from the anomalous and nuclear absorption is
11\%~\cite{Arnaldi:2007zz}. This is the dominating source of
systematic uncertainty in the analysis of the yield.

By multiplying the ratio in eq.~(\ref{eq:ratio}) by
eq.\eqref{eq:psi-vs-ncoll}, one obtains $\left<\phi\right>$ as a
function of the number of binary collisions $N_{coll}$. The values of
$N_{coll}$ and of $N_{part}$ for each centrality bin are obtained by
fitting the multiplicity distribution with a Glauber model, as
described in sec.~\ref{sec:CentrSelection} and summarised in
Tab.~\ref{tab:inin_centr_bins}. 

Processes having different scaling properties as a function of
centrality, such as the $\phi$ and the $J/\psi$, in general give rise
to different multiplicity distributions and this could introduce a
bias when using wide centrality bins.  The analysis was repeated with
a much finer centrality binning and this potential bias was found to
be negligible.

The multiplicity integrated in centrality is
$\langle\phi\rangle=1.411\pm0.026\pm0.14$.

\begin{table}[ptb]
  \caption{$\phi$ yield as a function of centrality. The bin by bin systematic 
    error is reported in the table. The measurements are also affected by a 13\% 
    systematic uncertainty independent of centrality.}
  \label{tab:yields}
  \centering
  \begin{tabular}{c r @{$\pm$} c @{$\pm$} l }
    \hline\noalign{\smallskip}
    $N_{part}$ & \multicolumn{3}{c}{$\langle\phi\rangle$} \\
    \noalign{\smallskip}\hline\noalign{\smallskip}
    14&     0.172 & 0.007 & 0.021\\   
    41&     0.690 & 0.025 & 0.074\\
    78&     1.48  & 0.04  & 0.15\\
    132&    3.10  & 0.08 & 0.2\\
    177&    5.0   & 0.17 & 0.2\\
    \noalign{\smallskip}\hline
    
  \end{tabular}
\end{table}

The robustness of the measurement is checked repeating the analysis
integrated in centrality, with a different method having largely
independent systematics. The $\phi$ cross section is determined with
the formula:
\begin{equation}
\sigma_{\phi}=\frac{N_{\phi}}{L \mathrm{A}_{\phi}\epsilon_{rec}^{\phi}\epsilon_{ev}\epsilon_{trig}\epsilon_{DAQ}\, BR^{\phi}_{\mu\mu}},
\label{eq:phi_absolute_yield}
\end{equation}
where $L$ is the integrated luminosity, $\epsilon_{ev}$ is the event
selection efficiency, $\epsilon_{trig}$ is the trigger efficiency,
$\epsilon_{DAQ}$ is the efficiency related to the DAQ live-time.  The
integrated luminosity is given by $L=N_{inc} N_{tgt}$, where $N_{inc}$
is the number of incident ions as measured by the ZDC and Beam Tracker
detectors, while $N_{tgt}$ is the effective number of indium atoms in
the target system~\cite{Lourenco:2007}. The latter is affected by a
systematic uncertainty of $10\%$, because the transverse profile of
the beam is wider than the targets (except for the first one which had
a radius of 0.6 cm).  The trigger efficiency was determined with
special runs taken without the dimuon trigger. It is 81\% and is
affected by a 6\% systematic error.

The average yield is then obtained dividing the cross section by the
fraction of total In-In inelastic cross section, corresponding to the
cut $N_{ch} > 4$, as estimated with the Glauber Monte Carlo (see
sec.~\ref{sec:CentrSelection}). The total inelastic cross section
calculated with the Glauber model is $\sigma_{inel}^{InIn} = 4.8$~b,
and has a 10\% uncertainty, mostly due to the uncertainties on the
nuclear density profiles.
The centrality integrated $\phi$ cross section is $\sigma_{\phi}= 6.83
\pm 0.08$~b. This leads to $\langle \phi \rangle =
1.73\pm0.02\pm0.27$, in agreement with the result based on the
standard method, taking into account only the unrelated systematic
errors in the two approaches.

The results as a function of centrality, obtained by taking the
average of the two methods weighted by the systematic error, are
summarised in Tab.~\ref{tab:yields}.  The ratio
$\left<\phi\right>/N_{part}$ is seen to increase with the collision
centrality indicating the presence of an enhancement as in other
collision systems.

\begin{figure}[tbp]
  \centering
  \includegraphics[width=0.48\textwidth,height=0.48\textwidth]{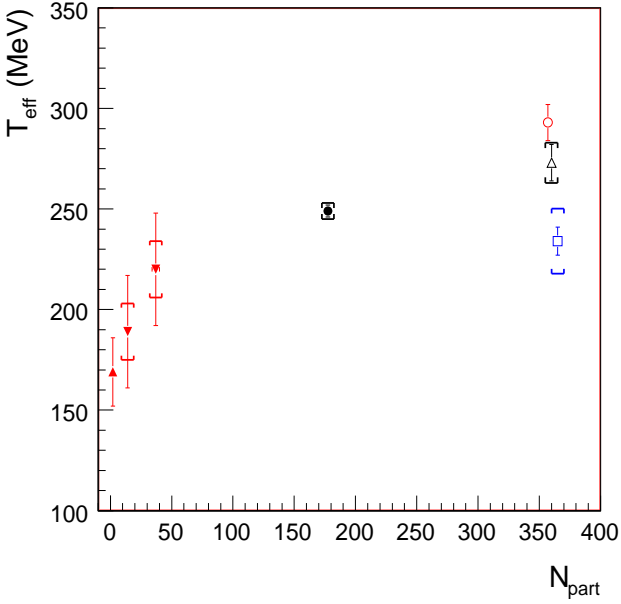}
  \caption{$T_{eff}$ for central collisions in
    several collision systems at 158~AGeV.  Full circle:
    $\phi\to\mu\mu$ in In-In (NA60); full upward triangle: $\phi\to
    K^+K^-$ in p-p (NA49); full downward triangle: $\phi\to K^+K^-$
    in C-C and Si-Si (NA49); open circle: $\phi\to K^+K^-$ in Pb-Pb
    (NA49); open square: $\phi\to\mu\mu$ in Pb-Pb (NA50); open
    triangle: $\phi\to K^+K^-$ in Pb-Pb (CERES).}
  \label{fig:T_vs_npart_central}
\end{figure}
\begin{figure}[t]
  \centering
  \includegraphics[width=0.48\textwidth,height=0.48\textwidth]{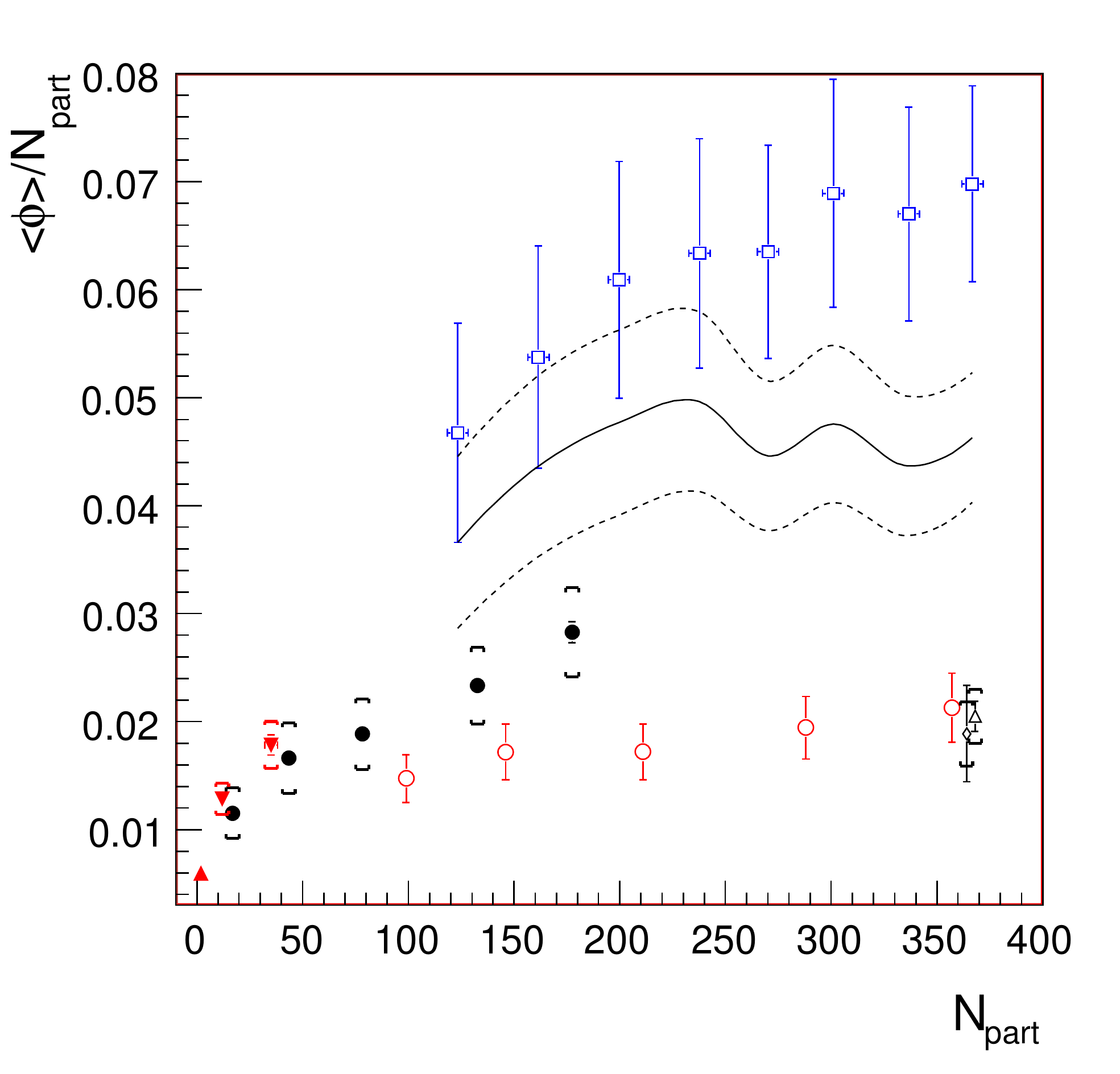}
  \caption{$\langle\phi\rangle/N_{part}$ in full phase space in
    several collision systems at 158~AGeV.  Full circles:
    $\phi\to\mu\mu$ in In-In (NA60); full upward triangle: $\phi\to
    K^+K^-$ in p-p (NA49); full downward triangles: $\phi\to K^+K^-$
    in C-C and Si-Si (NA49); open circles: $\phi\to K^+K^-$ in Pb-Pb
    (NA49); open squares: $\phi\to\mu\mu$ in Pb-Pb (NA50); open
    triangle: $\phi\to K^+K^-$ in Pb-Pb (CERES); open diamond:
    $\phi\to ee$ in Pb-Pb (CERES).  Continuous line: NA50 points
    extrapolated to full phase space with $T_{eff}$ as measured by
    NA49, dashed lines: uncertainty on these points. Square brackets:
    systematic error (the error bar of the Pb-Pb data includes
    statistical and systematic error).  C, Si and peripheral In points
    slightly displaced to improve readability.}
  \label{fig:yield}
\end{figure}

\subsection{Discussion of the results on $T_{eff}$ and yield}
\label{sec:discussion}

As mentioned in the introduction, the discrepancies in absolute yields
and $T_{eff}$ values observed by NA49 and NA50 became historically
known as the \emph{\f~puzzle}~\cite{Shuryak:1999zh,Rohrich:2001qi}. To
explain those differences, it was originally proposed that in-medium
effects and kaon absorption or rescattering could prevent the
reconstruction of in-matter $\phi\to K\bar K$ decays, in particular at
low transverse momentum, while the $\phi$ mesons decaying in the
lepton channel would not be affected.  This would lead to reduced
yields and enhanced $T_{eff}$ values in the $K\bar{K}$ channel as
compared to the $\mu\mu$ channel.  In this section, all the
measurements performed at the SPS, including the new one in In-In, are
compared and the hypothesis of the physical mechanism described above
is critically reviewed, considering the facts in favour of and against
it.

Fig.~\ref{fig:T_slope_na49na50range} shows the NA60 $T_{eff}$
obtained performing the fits in the $p_T$ windows corresponding to the
coverage of NA49 (left) and NA50 (right).
As mentioned in sec.~\ref{sec:transv-moment-distr}, the difference of
the temperature in the two $p_T$ windows in In-In is around 10 MeV,
indicating a modest radial flow effect.  Although a similar pattern is
observed in Pb-Pb collisions -- a stronger increase at low $p_T$
(NA49) and a flattening at high $p_T$ (NA50) -- the maximum difference
is, for the most central Pb-Pb collisions, about 70 MeV.  This
difference seems not to be consistent with the $T_{eff}$ variation
implied by radial flow as observed in the NA60 data.

In the common fit range, the NA60 $T_{eff}$ values in In-In are $\sim10$ MeV
larger than those measured by NA50,
while there is a seeming agreement between NA60 and NA49 in the common
fit range. However, the NA49 errors are large in this region and, in
addition, a direct comparison of the  values of $T_{eff}$
between NA60 and the Pb-Pb experiments for the same $N_{part}$ is not
straightforward, due to the different geometry of the collision (the
relative weight of the thin nuclear halo is higher in peripheral Pb-Pb
collision than in a smaller system).

The temperature variation in different collision systems, when
restricting to central collisions only, is shown in
Fig.~\ref{fig:T_vs_npart_central}. This picture includes also the NA49
results for C-C, Si-Si and the CERES result in the kaon channel for
central Pb-Pb (the dielectron measurement has a very large error).
The NA49 and CERES Pb-Pb points are above the In-In point by
$\gtrsim30$ MeV. Part of this difference can be related to radial
flow, since in Pb-Pb a larger lifetime of the fireball would lead to a
larger effect in central collisions.  However, the $T_{eff}$ measured
for central Pb-Pb collisions is $\sim 290$~MeV, very close to that of
the proton~\cite{Alt:2006dk}. The blast-wave analysis of the NA60
data, on the other hand, shows a clear hierarchy in the freeze-out,
with the $\rho$ having a maximal flow due to its coupling to pions and
the $\phi$ being the least coupled meson
(Fig.~\ref{fig:blast_wave}). The temperature of the freeze-out
component of the $\rho$ in In-In reaches 300 MeV or more, 50 MeV above
the $\phi$. One would therefore expect an even larger difference in
Pb-Pb collisions between the $T_{eff}$ of the $\phi$ and of the
proton, which is strongly coupled to the pions via the $\Delta$
resonance and with a mass larger than the $\rho$. This suggests that
the high $T_{eff}$ seen by NA49 and CERES in the kaon channel cannot
be ascribed entirely to radial flow.

\begin{figure*}[tbp]
  \centering
  \includegraphics[width=0.48\textwidth,height=0.48\textwidth]{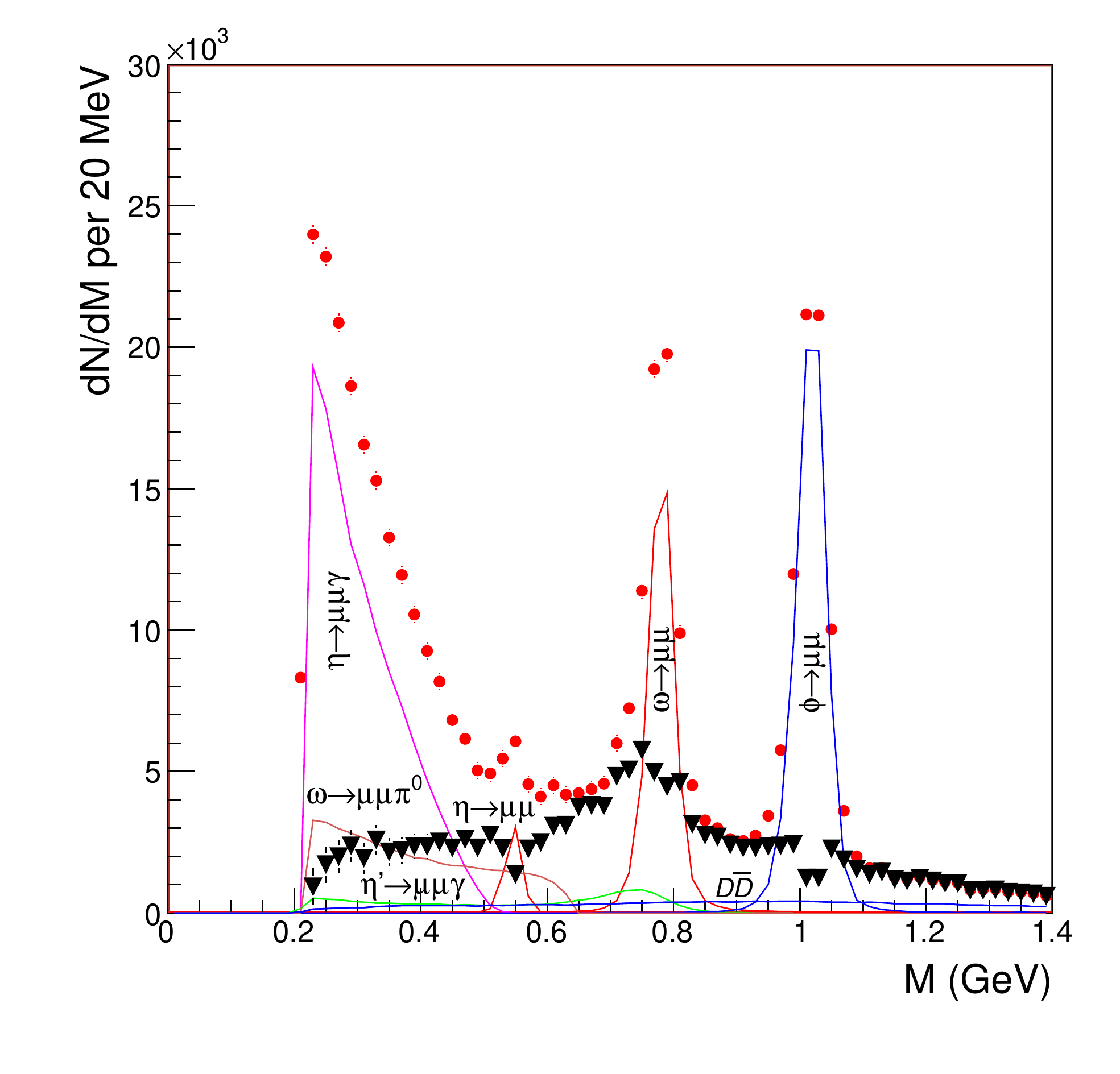}
  \includegraphics[width=0.48\textwidth,height=0.48\textwidth]{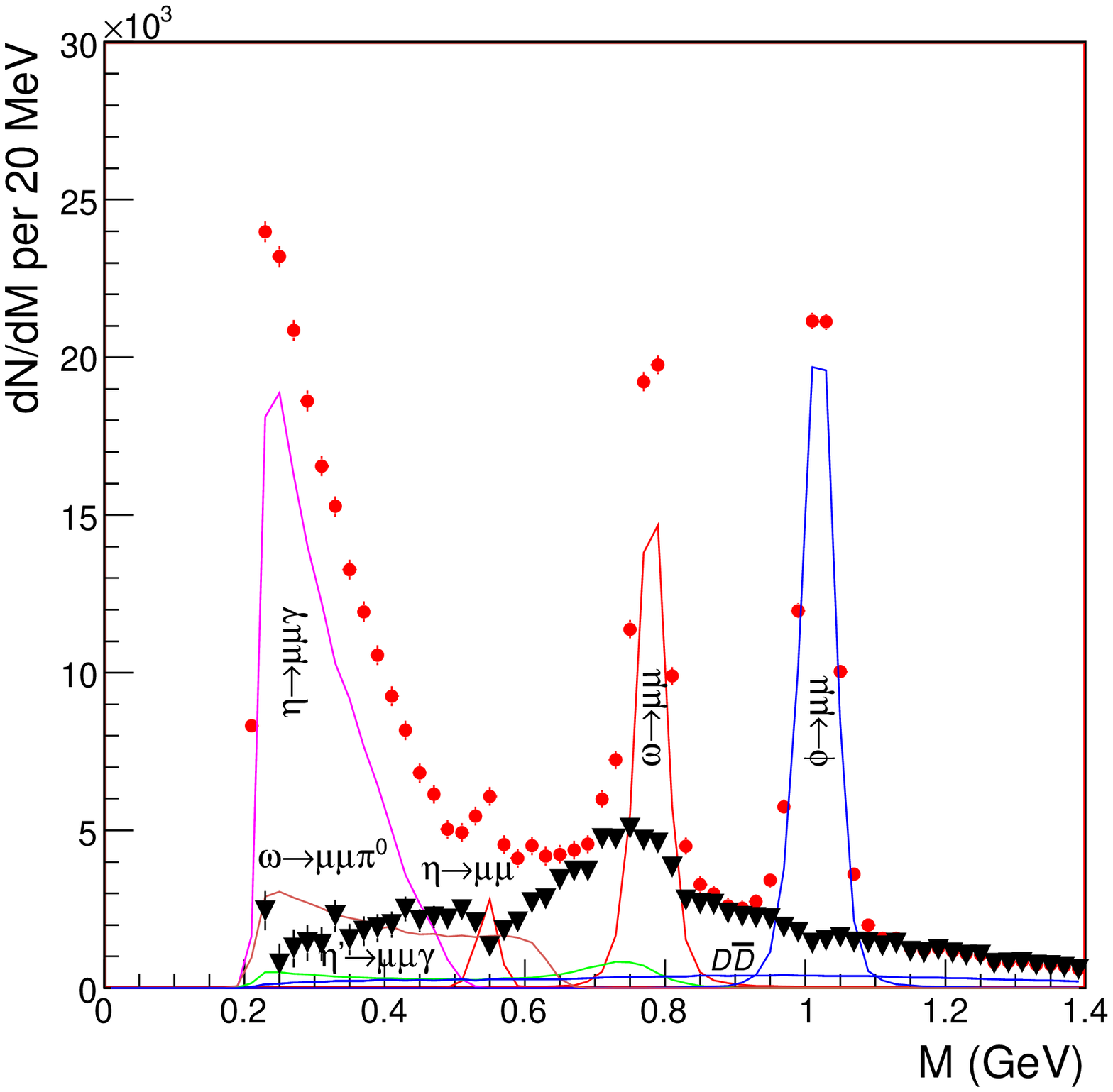}
  \caption{Excess mass spectrum for data integrated in centrality,
    obtained without (left) and with (right) the Monte Carlo fine
    tuning. Red circles: data; black triangles: excess data; lines:
    individual processes. }
  \label{fig:excess}
\end{figure*}

The $\phi$ enhancement in In-In is compared to previous SPS
measurements~\cite{Friese:2002re,Alt:2004wc,Alessandro:2003gy,Adamova:2005jr}
in Fig.~\ref{fig:yield}.

An unambiguous comparison to the NA50 result in full phase space is
not possible, due to lack of consensus on the value of the $T_{eff}$
parameter in Pb-Pb collisions, but in general the NA50 enhancement is
considerably higher than what is seen in In-In at any centrality.  An
extrapolation to full $p_{T}$ with the NA50 $T_{eff}$ values
($T\sim220-230$ MeV independent of centrality) leads to values higher
by a factor 2 or more with respect to In-In.  Even extrapolating with
the NA49 $T_{eff}$ values ($T\sim250-300$ MeV) would lead to values
significantly higher (as shown by the continuous line in
Fig.~\ref{fig:yield}, the dashed lines represent the propagated
uncertainty on these values).

The NA49 measurements\footnote{NA49 has characterized centrality of
  the collision using 2 quantities: the ``number of wounded nucleons''
  and the ``number of participants''. The one directly comparable to
  our $N_{\mathrm{part}}$ is the number of wounded nucleons. The
  values for the different Pb-Pb bins are taken from
  Ref.~\cite{Alt:2006dk}.} show that the $\langle\phi\rangle$ breaks
$N_{part}$ scaling and that $\left<\phi\right>/N_{part}$, for central
collisions as a function of the colliding system, already saturates
for Si-Si.  The $\phi$ enhancement in In-In is closer to NA49 than
NA50, but it is not in agreement either. The two In-In most peripheral
points are in agreement with C-C and Si-Si. However, for increasing
centrality there is no saturation in In-In, while the NA49 Pb-Pb
points show a flatter trend, which remains consistently smaller at any
centrality. One could ask whether the increase of the yield ratio
NA60/NA49 as a function of centrality could be ascribed to a
suppression mechanism present in Pb-Pb that tends to decrease the
overall enhancement. Restricting the comparison among In-In and Pb-Pb
central collisions, the dimuon NA60 measurement remains larger than
the kaon Pb-Pb points.

Summarizing, the large NA49/NA50 difference in central collisions for
$T_{eff}$ (70 MeV) and the yield (a factor 4), was hardly accounted for by
theoretical models.  A difference, though sm\-al\-ler, is also present
in central collisions in $T_{eff}$ and yield, when considering the
NA60 muon measurement. This difference remains difficult to explain
and hints for the presence of a physical effect.  Whether this is
present also in In-In, or peculiar only to Pb-Pb, would require the
comparison to $\phi\to K\bar K$ in In-In. Preliminary NA60 results of
$\phi\to K^+ K^-$ suggest that there are no differences within the
errors among the two channels in In-In~\cite{adf-qm09}.

One has to notice that the CERES dielectron and kaon points for
central Pb-Pb are in agreement and in addition, both the NA50
$T_{eff}$ and yield do not agree with NA60 either. Thus, at the moment
it is not possible to include into a coherent picture all the
measurements performed so far in the different collision systems.

\subsection{Mass and width}
\label{sec:Mass_Width}

\begin{figure*}[tpb]
  \centering
  \includegraphics[width=0.48\textwidth,height=0.48\textwidth]{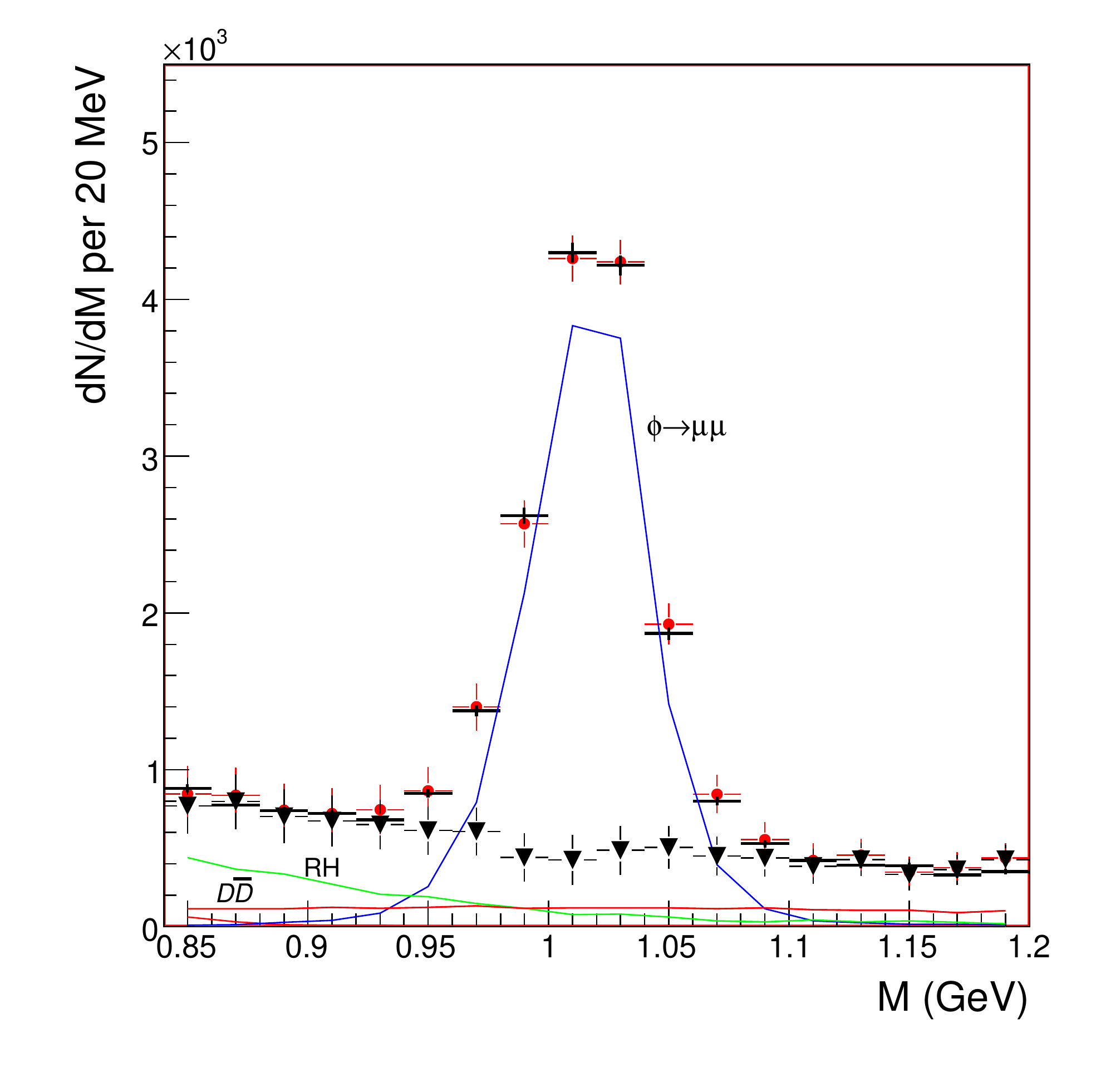}
  \includegraphics[width=0.48\textwidth,height=0.48\textwidth]{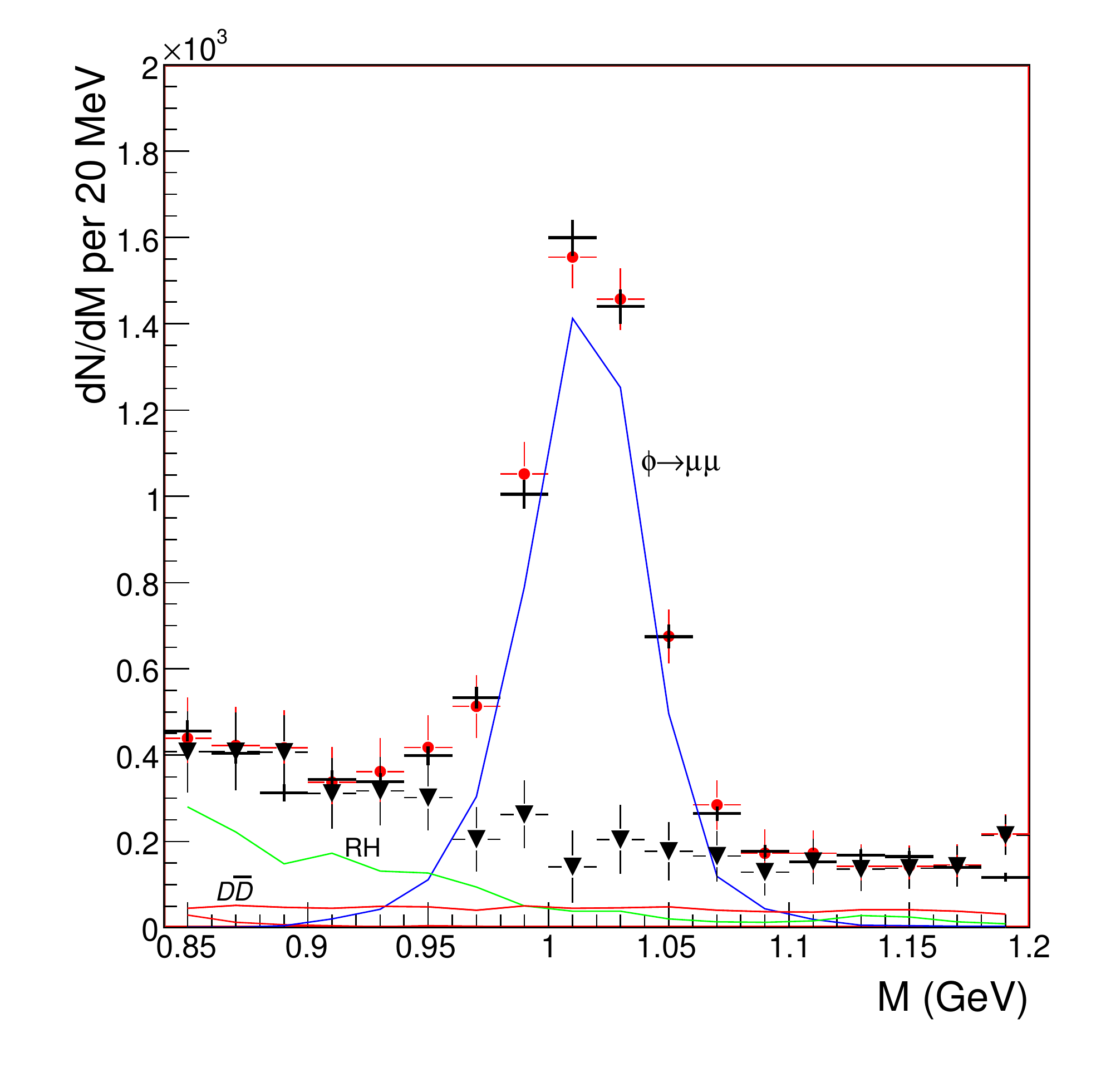}
  \caption{\f~mass region in the two most central bins for $p_{T} <
    0.6~\mathrm{GeV}$ (left) and all centralities, $p_{T} <
    0.2~\mathrm{GeV}$ (right). Red circles: data; black triangles:
    excess data; lines: individual processes (RH: model of
    Ref.~\cite{vanHees:2007th}); black crosses: result of fit
    performed using the Monte Carlo shapes.}
  \label{fig:excess_phi_peak}

\end{figure*}
 
As mentioned in the introduction, the spectral function of the \f~
could be modified in the medium, resulting in changes of its mass and
width.  The lifetime of the $\phi$ is $44$~fm, larger than the
estimated fireball lifetime ($\sim7$~fm). Any medium modified
component would therefore sit under a large unmodified peak produced
after freeze-out. In order to distinguish a possible in-medium
component from instrumental effects, a careful tuning of the Monte
Carlo simulation is required. This was not strict\-ly necessary in
previous analysis of the low mass region
excess~\cite{Damjanovic:2008ta,Damjanovic:2006}, as this structure,
mostly due to $\pi\pi\to\rho^{*}\to\mu\mu$, represents a large
fraction of the total yield (the lifetime of the $\rho$ is $\sim
1.3$~fm).

Uncertainties in the detectors alignment, materials, energy loss
compensation and precision of the magnetic field maps lead to residual
differences between the Monte Carlo and the real data. These were
compensated by fine-tuning the Monte Carlo against the peripheral
data, where no in-medium effects are expected. The tuning was
performed considering the whole mass spectrum below 1 GeV and required
a global $0.3\%\pm0.1\%$ mass shift and a $4\%\pm1\%$ change in the
mass resolution (previous NA60 analyses only included the mass shift
correction). The uncertainties in this tuning are the main source of
systematic error for the results discussed in this section.

In order to look for an in-medium component around the \f~peak, the
excess was isolated using the same subtraction procedure already
extensively used for the low mass region~\cite{Damjanovic:2006}. The
$\omega$, $\phi$ and $\eta$ Dalitz were subtracted from the total data
using the Monte Carlo shapes and fixing the yield using solely local
criteria; the $\omega$ Dalitz and the $\eta$ two body were then
subtracted by binding them to these processes via branching ratios.
The quality of the subtraction can be judged in Fig.~\ref{fig:excess}
which shows, for data integrated over centrality, the excess before
and after the fine tuning. While the tuning is essentially constrained
by the narrow $\phi$ peak, it improves the subtraction over the whole
mass spectrum: the evident artifacts visible in the left panel around
all the narrow resonances ($\eta$, $\omega$ and $\phi$) disappear
after the tuning.  The continuum left after the subtraction under the
\f~peak is smooth: no evidence for a medium modified $\phi$ can be
seen in this figure.  Indeed, an in-medium broadened component would
manifest itself as broader tails extending to the sides. In this case,
the subtraction of the vacuum component should not leave a smooth
continuum, but rather a broad structure or a dip, which is not
observed.

As mentioned in the introduction, the KEK-PS E325 experiment reported
evidence for a large excess to the left of $\phi$ peak in p-Cu at 12
GeV for $\beta \gamma < 1.25$, corresponding to a produced $\phi$
moving very slowly inside the heavy target nucleus at rest. This
situation is not directly comparable to the high energy regime
discussed in this paper.  Here, a fireball is produced with particle
momenta boosted in the forward direction, therefore the relevant
variable to select $\phi$ mesons mostly sensing the medium is
transverse momentum.  Moreover, in a high energy heavy ion collision,
as opposed to p-A, the $\phi$ can be produced not only at chemical
freeze-out, but also in the hot hadronic medium via kaon reactions.
Fig.~\ref{fig:excess_phi_peak} shows the region around the $\phi$ mass
peak for different $p_T$ and centrality selections: the left panel
shows the two most central bins and for $p_T<0.6$ GeV, while the right
panel shows the data integrated in centrality for $p_T<0.2$ GeV. The
continuum from the subtraction is smooth also in this case.
Furthermore, a fit was performed using the Monte Carlo shapes for the
different processes. The continuum was described with the model of
Ref.\cite{vanHees:2007th}, which includes, besides the in-medium $\rho$,
multi-pion and partonic processes relevant in this mass region. The
model of Ref.~\cite{Ruppert:2007cr}, having different weights for the
multi-pion and partonic processes, does not change the conclusions,
since it produces a similar yield below the \f~peak. The $\chi^2/ndf$
is $\sim1$ in all cases. There is no evidence of any mass shift or
broadening, even at high centrality and low $p_T$, where one would
expect to observe the strongest in-medium effects.
\begin{figure*}[tbp]
  \centering
  \includegraphics[width=0.48\textwidth,height=0.48\textwidth]{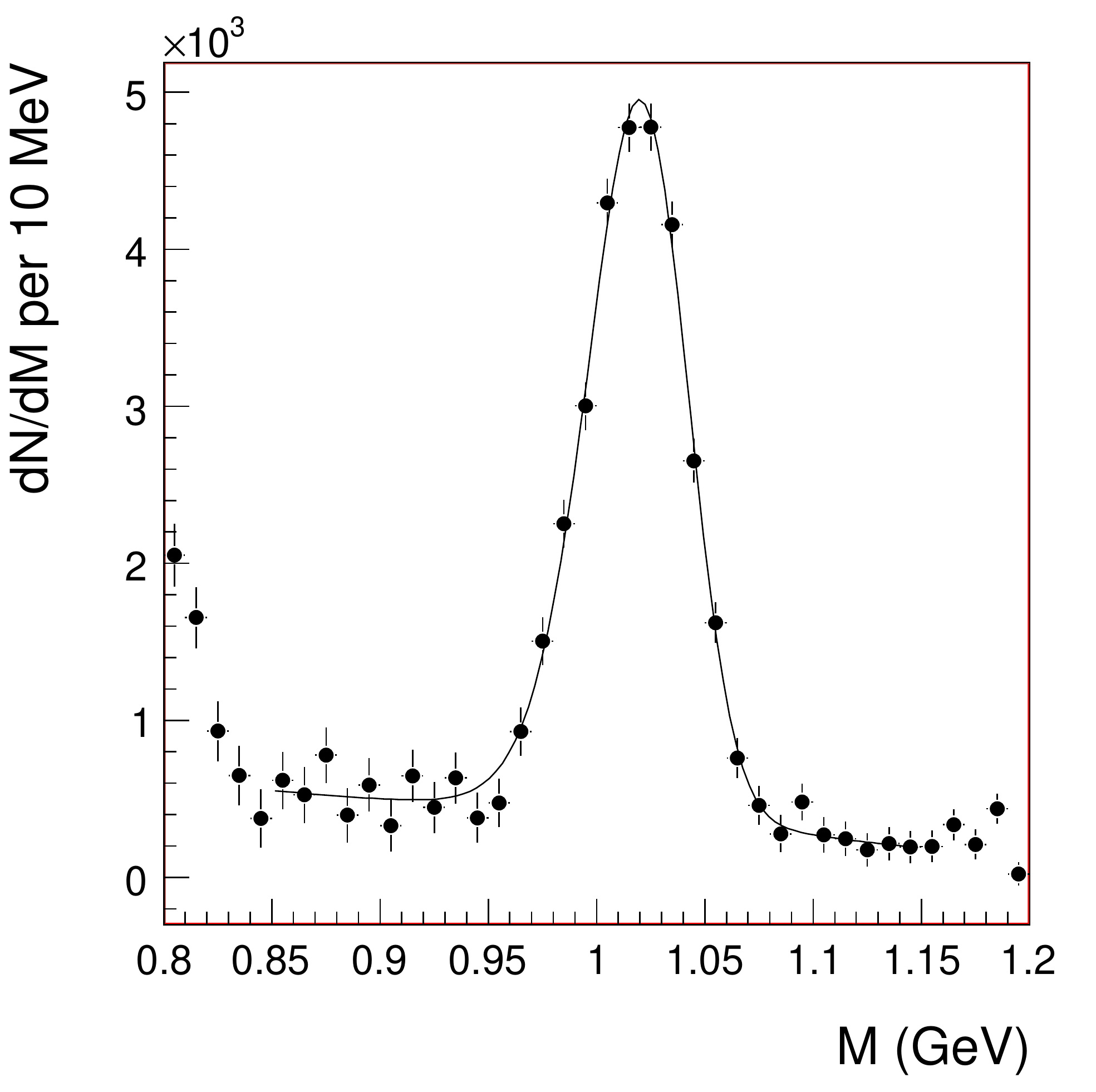}
  \includegraphics[width=0.48\textwidth,height=0.48\textwidth]{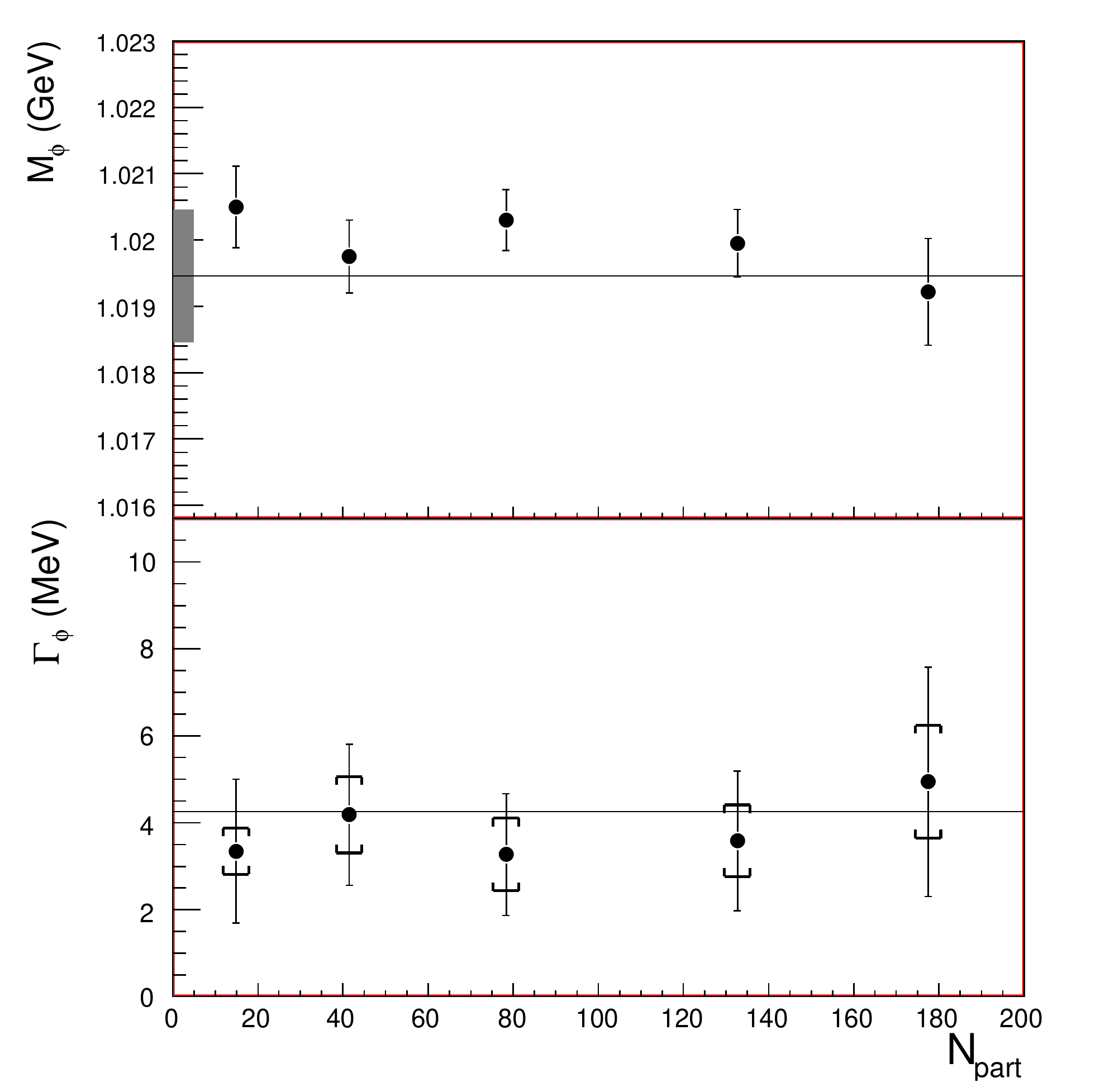}

  \caption{Left panel: fit of the $\phi$ mass peak, semi-central bin
    ($N_{part} = 133$) with a relativistic Breit-Wigner convoluted
    with the detector response, on top of a linear background. Right
    panel: mass (top) and width (bottom) of the \f~as a function of
    centrality. Shaded box: systematic uncertainty common to all
    centrality bins.}
  \label{fig:bw-fits}
\end{figure*}

In order to estimate quantitatively the pole mass and total width of
the $\phi$ peak, data were fitted with the convolution of a
relativistic Breit-Wigner and of a function parametrizing the detector
response, on top of a linear background. In a first step, the
parameters of the detector response were adjusted on the Monte Carlo
simulation. The data are then fitted leaving the mass and width of the
Breit-Wigner as free parameters. The quality of the fit is very good,
as shown in the left panel of Fig.~\ref{fig:bw-fits}, corresponding to
the semi-central bin, $N_{part} = 133$. The results the mass and width
as a function of centrality are shown in the right panel of
Fig.~\ref{fig:bw-fits}.  The systematic error was estimated changing
the fit range, the background function and propagating the
uncertainties in the Monte Carlo tuning.  The mass and width are
compatible with the PDG values and independent of centrality, although
our sensitivity on the width is limited by the mass resolution, which
is much larger than the natural width (this is reflected in the large
statistical error in the right panel of Fig.~\ref{fig:bw-fits}).

While the data can be described by a vacuum $\phi$, we set an upper
limit on a modified component using the only existing prediction for
in-medium production of $\phi$ in 158~AGeV In-In
collisions~\cite{vanHees:2007th}. This corresponds to an in-medium
spectral function folded over the phase-space fireball evolution up to
freeze-out and appears significantly broadened ($> 80$ MeV), so that
the data should have sufficient sensitivity - in terms of mass
resolution - to discriminate it. The region around the $\phi$ peak was
fitted adding an in-medium $\phi$ component to the unmodified one. In
the centrality integrated data, the fraction in-medium/vacuum $\phi$
is compatible with zero within one sigma with an upper limit of $0.08
\pm 0.03~(\mathrm{syst})$ at 95\% confidence level. In the two most
central bins the fraction remains compatible with zero within one
sigma and the upper limit increases to $0.15 \pm
0.03~(\mathrm{syst})$, due the larger statistical uncertainty. The
fraction predicted by Ref.~\cite{vanHees:2007th} for semi-central data
(roughly equivalent to centrality integrated data) is $\sim 4$\%. The
limitation in statistics prevents us from setting a meaningful upper
limit at high centrality and low $p_{T}$.


\section{Conclusions and outlook}
\label{sec:conclusions}

In this paper a complete characterization of $\phi\to\mu\mu$
production in In-In collisions has been provided by measuring the
differential spectra, yield,  mass and width.

The rapidity distribution width does not depend on centrality within
errors, and is found to be compatible with previous SPS measurements.
The decay angle distributions show no evidence for polarization in any
of the reference frames used, independent of centrality and
$p_T$. This confirms the lack of anisotropy previously reported by
NA60. This is a necessary though not sufficient condition for particle
production from a thermalized medium. Indeed, the centrality
dependence suggest that the polarization is zero or very close to zero
also in peripheral collisions, close to the limit of p-p like
interactions.

The study of the transverse momentum distributions shows a small
dependence of the $T_{eff}$ parameter on the $p_T$ window considered,
to be ascribed presumably to radial flow. A blast wave analysis shows
a clear hierarchy of the freeze-out, with the $\phi$ being the least
coupled particle to the medium.  The difference between $T_{eff}$
measured in Pb-Pb collisions by NA49 and NA50 is much larger and
cannot be explained by the different $p_T$ windows explored.  The
$T_{eff}$ value for most central collisions in In-In remains
significantly lower than the kaon point in Pb-Pb measured by NA49 and
CERES.

The yield measured by NA50 at high $p_T$, when extrapolated to full
$p_T$, exceeds the NA60 data points for the same number of
participants and seems to lie outside of the general systematics.  The
NA60 yield, on the other hand, remains consistently larger than the
NA49 and CERES Pb-Pb measurement.

The new measurements in $T_{eff}$ and yield, when compared to the kaon
channel in Pb-Pb, suggest that there may be a physical effect leading
to a difference in the two channels.  Further clarification may come
from the NA60 measurement of the $\phi\to K^+K^-$ channel in In-In
collisions, which is currently in progress and will be reported in a
forthcoming paper.  At the moment, it is not possible to include all
the existing SPS measurements in a coherent framework.

The mass and width were studied as a function of centrality and
transverse momentum. No evidence for in-medium modifications was
found.  Our data were also compared to the only existing prediction
for in-medium $\phi$, produced in In-In collisions at 158 AGeV. The
yield of such a component is compatible with zero, with an upper limit
of about 8\% for the centrality integrated data.

\section*{Acknowledgments}

We acknowledge support from the BMBF (Heidelberg
group) as well as from the C. Gulbenkian Foundation
and the Swiss Fund Kidagan (YerPHI group).


\bibliographystyle{epjc} 
\bibliography{na60_phi_inin_epj}

\end{document}